\documentclass[twocolumn,times]{aastex6}

\usepackage{ulem}
\usepackage{graphicx}
\usepackage{amsmath}
\usepackage{epsfig} 
\usepackage{enumitem}

\newcommand{\RA}[3]{#1^\mathrm{h}#2^\mathrm{m}#3^\mathrm{s}}
\newcommand{\DEC}[3]{#1^{\circ}#2'#3''}

\newcommand{\kms}{km~s{$^{-1}$}}
\newcommand{\Vlsr}{V{$_{LSR}$}}

\newcommand{\hh}{H{$_2$}}

\newcommand{\Tco}{{$^{12}$CO}}

\newcommand{\Lsol}{L$_{\odot}$}
\newcommand{\Msol}{M$_{\odot}$}

\newcommand{\HST}{{\it HST}}

\begin{document}

\title{SPIRITS:  Uncovering Unusual Infrared Transients With {\it Spitzer}} %A Systematic Mid-Infrared Transient Search \\ %{\it Experiment Design, Software Pipelines and First Discoveries}}
\author{Mansi M. Kasliwal\altaffilmark{1},
John Bally\altaffilmark{2},
Frank Masci \altaffilmark{3},
Ann Marie Cody\altaffilmark{4},
Howard E. Bond\altaffilmark{7,10},
Jacob E. Jencson\altaffilmark{1},
Samaporn Tinyanont \altaffilmark{1},
Yi Cao \altaffilmark{1},
Carlos Contreras\altaffilmark{5},
Devin A. Dykhoff\altaffilmark{6},
Samuel Amodeo\altaffilmark{6},
Lee Armus \altaffilmark{3},
Martha Boyer\altaffilmark{8,22},
Matteo Cantiello\altaffilmark{9},
Robert L. Carlon\altaffilmark{6},
Alexander C. Cass\altaffilmark{6},
David Cook\altaffilmark{1},
David T. Corgan\altaffilmark{6},
Joseph Faella\altaffilmark{6},
Ori D. Fox\altaffilmark{10},
Wayne Green\altaffilmark{2},
Robert Gehrz\altaffilmark{6},
George Helou \altaffilmark{3},
Eric Hsiao\altaffilmark{11},
Joel Johansson\altaffilmark{12},
Rubab M. Khan\altaffilmark{8},
Ryan M. Lau \altaffilmark{1,20},
Norbert Langer\altaffilmark{13},
Emily Levesque\altaffilmark{14},
Peter Milne\altaffilmark{15},
Shazrene Mohamed\altaffilmark{16,21,23},
Nidia Morrell\altaffilmark{5},
Andy Monson\altaffilmark{7},
Anna Moore\altaffilmark{1}, 
Eran O. Ofek\altaffilmark{12},
Donal O' Sullivan\altaffilmark{1},
Mudumba Parthasarthy\altaffilmark{17},
Andres Perez\altaffilmark{1},
Daniel A. Perley\altaffilmark{18},
Mark Phillips\altaffilmark{5},
Thomas A. Prince \altaffilmark{1},
Dinesh Shenoy\altaffilmark{6},
Nathan Smith\altaffilmark{15},
Jason Surace\altaffilmark{19},
Schuyler D. Van Dyk\altaffilmark{3},
Patricia Whitelock\altaffilmark{16,21},
Robert Williams\altaffilmark{10}
}

\altaffiltext{1}{Division of Physics, Mathematics and Astronomy, California Institute of Technology, Pasadena, CA 91125, USA}
\altaffiltext{2}{Center for Astrophysics and Space Astronomy, University of Colorado, 389 UCB, Boulder, CO 80309, USA}
\altaffiltext{3}{Infrared Processing and Analysis Center, California Institute of Technology, Pasadena, CA 91125, USA}
\altaffiltext{4}{NASA Ames Research Center, Moffett Field, CA 94035, USA}
\altaffiltext{5}{Las Campanas Observatory, Carnegie Observatories, Casilla 601, La Serena, Chile}
\altaffiltext{6}{Minnesota Institute for Astrophysics, School of Physics and Astronomy, 116 Church Street, S. E., University of Minnesota, Minneapolis, MN 55455, USA}
\altaffiltext{7}{Dept. of Astronomy \& Astrophysics, Pennsylvania State University, University Park, PA 16802 USA}
\altaffiltext{8}{NASA Goddard Space Flight Center, MC 665, 8800 Greenbelt Road, Greenbelt, MD 20771 USA} %Observational Cosmology Lab, Code 665
\altaffiltext{9}{Kavli Institute for Theoretical Physics, University of California, Santa Barbara, CA 93106, USA}
\altaffiltext{10}{Space Telescope Science Institute, 3700 San Martin Dr., Baltimore, MD 21218 USA}
\altaffiltext{11}{Department of Physics, Florida State University, 77 Chieftain Way, Tallahassee, FL, 32306, USA}
\altaffiltext{12}{Benoziyo Center for Astrophysics, Weizmann Institute of Science, 76100 Rehovot, Israel}
\altaffiltext{13}{Argelander-Institut für Astronomie Auf dem Hügel 71. D-53121 Bonn, Germany}
\altaffiltext{14}{Department of Astronomy, University of Washington Seattle, Box 351580, WA 98195-1580 USA}
\altaffiltext{15}{Steward Observatory, University of Arizona, Tuscon, AZ 85721, USA}
\altaffiltext{16}{SAAO, PO Box 9, Observatory, 7935, South Africa}
\altaffiltext{17}{Indian Institute of Astrophysics, Koramangala,  Bangalore 560034, India}
\altaffiltext{18}{Dark Cosmology Centre, Niels Bohr Institute, Juliane Maries Vej 30, Copenhagen AF, DK-2100, Denmark}
\altaffiltext{19}{Eureka Scientific, Inc. 2452 Delmer Street Suite 100 Oakland, CA 94602-3017 USA}
\altaffiltext{20}{Jet Propulsion Laboratory, California Institute of Technology, 4800 Oak Grove Drive, Pasadena, CA 91109, USA}
\altaffiltext{21}{Astronomy Department, University of Cape Town, University of Cape Town, 7701, Rondebosch, South Africa}
\altaffiltext{22}{Department of Astronomy, University of Maryland, College Park, MD 20742 USA}
\altaffiltext{23}{National Institute for Theoretical Physics, Private Bag X1, Matieland, 7602, South Africa}
%\altaffiltext{}{}

\begin{abstract}
We present an ongoing, systematic search for extragalactic infrared transients, dubbed SPIRITS ---
SPitzer InfraRed Intensive Transients Survey. In the first year, using {\it Spitzer}/IRAC, we searched 190 nearby galaxies %to a depth of 20\,mag 
with cadence baselines of one month and six months. We discovered over 1958 variables and 43 transients. 
Here, we describe the survey design and highlight 14 unusual infrared transients with no optical counterparts to deep limits, which
we refer to as SPRITEs (eSPecially Red Intermediate Luminosity Transient Events).  SPRITEs are in the infrared luminosity gap between 
novae and supernovae, with [4.5] absolute magnitudes between $-$11 and $-$14 (Vega-mag) and [3.6]-[4.5] colors 
between 0.3\,mag and 1.6\,mag. The photometric evolution of SPRITEs is diverse, ranging from $<$0.1 mag yr$^{-1}$ to $>$7\,mag yr$^{-1}$. 
SPRITEs occur in star-forming galaxies. We present an in-depth study of one of them, SPIRITS\,14ajc  in Messier\,83, which shows shock-excited molecular hydrogen emission.
%suggesting that it represents the birth of a massive star binary driving a shock through a molecular cloud.  
This shock may have been triggered by the dynamic decay of a non-hierarchical system of massive stars
that led to either the formation of a binary or a proto-stellar merger. 
\end{abstract}


\section{Introduction}
The systematic study of explosive transients and eruptive variables is growing by leaps and bounds
especially with the advent of wide-field synoptic imaging. 
Recently, multiple new classes of optical transients (e.g., \citealt{Kasliwal2012}) and new radio transients (e.g., \citealt{Thornton2013}) 
have been uncovered. Yet, the dynamic infrared (IR) sky is hitherto largely unexplored.

While the optical is a powerful band to explore supernovae and novae, it is blind to 
transients and eruptive variables that are either self-obscured or located in dusty regions
(e.g., molecular clouds). IR follow-up of optically discovered transients shows that
IR emission dominates in supernovae with circumstellar interaction, particularly at late-time \citep{Fox2011,Fox2013}. 
We are now aware of at least two new classes of explosive transients where the bulk of the 
emission is in the IR -- stellar mergers (associated with luminous red novae, e.g. V1309 Sco; \citealt{Tylenda2011}) and 
electron capture supernovae  (eCSNe; associated with intermediate luminosity red transients, e.g. NGC300-OT; \citealt{Bond2009}, 
SN\,2008S; \citealt{Prieto2008}).  

Some efforts have been undertaken to look for IR transients and variables with the {\it Spitzer} Space Telescope \citep{Werner2004,Gehrz2007}.
A blind search for IR transients in repeated imaging of the Bootes field revealed a superluminous supernova \citep{Kozlowski2010}.
Searches targeting nearby star forming regions in the Milky Way have shown a plethora of young star variability \citep{Cody2014,Rebull2014}. 
Searches for variable, obscured asymptotic giant branch stars in nearby galaxies is being undertaken by the DUSTiNGS survey \citep{Boyer2015}.
Searches for obscured supernovae in starburst galaxies, using  {\it Spitzer} \citep{Fox2012}, {\it HST} \citep{Cresci2007} and high-resolution
ground-based adaptive optics imaging (e.g., \citealt{Mattila2007}), have revealed a few candidates \citep{Kankare2008,Kankare2012}. 

Motivated thus, we began a systematic search for mid-IR transients in 
nearby galaxies with {\it Spitzer}.  Here, we present the experiment design (\S\ref{sec:experiment}), 
the software pipelines (\S\ref{sec:software}), the discoveries in the first year (\S\ref{sec:discoveries})
and a case-study (\S\ref{sec:14ajc}). We conclude with reflections on a possible way forward to chart 
the dynamic IR sky (\S\ref{sec:conclusion}). 

\section{Experiment Design}
\label{sec:experiment}

\subsection{Galaxy Sample, Cadence and Depth}

The SPIRITS experiment uses the IRAC instrument (FoV 5'$\times$5'; \citealt{Fazio2004}) aboard 
the warm {\it Spitzer} telescope to search for IR transients at 3.6 $\mu$m  ([3.6]) and 4.5 $\mu$m ([4.5]). 
This is a search targeting 190 nearby galaxies selected using three criteria:
(i) 37 galaxies out to 5\,Mpc spanning diverse galaxy 
environments: early-type galaxies, late-type galaxies, dwarf galaxies and giant galaxies 
%(note: some galaxies had to be excluded as they were too large on sky to efficiently map with IRAC). 
(ii) 116 luminous galaxies between 5\,Mpc and 15\,Mpc. Our sample of galaxies captures a total of 
2.0$\times$10$^{12}$ L$_{\odot}$ in the B-band within 15 Mpc. This is 83\% of the total B-band starlight within the $<$15 Mpc volume.
(iii) The 37 most luminous and most massive galaxies in the Virgo Cluster (17\,Mpc). These galaxies total
 1.6$\times$\,10$^{12}$\,L$_{\odot}$ i.e. 66\% of the total B-band luminosity of the cluster. Furthermore, these 
 galaxies total 9.1\,$\times$\,10$^{11}$\,M$_{\odot}$ i.e. 71\% of the total stellar mass of the cluster. 

Tables summarizing properties of galaxies in this sample in detail is presented in \citealt{SPIRITS10}. 
In 2015 and 2016, we added star-forming regions in galaxies too large to map with IRAC otherwise \citep{SPIRITS11}. 
In 2017 and 2018, we down-sized the sample to focus on the most luminous and most massive galaxies \citep{SPIRITS13}.  

In 2014, each of these galaxies was imaged three times by SPIRITS, with cadence baselines of 1 month
and 6 months. In 2015 and 2016, additional shorter cadence baselines of 1 week and 3 weeks were added. 
Archival data provides us additional multi-year baselines. A histogram of cadence baselines is available in  \citealt{SPIRITS11}.

Each SPIRITS pointing is seven dithered 100\,s exposures in each IRAC filter. The limiting magnitude (as defined for a 5$\sigma$ point source Vega magnitude) 
in each SPIRITS epoch is 20\,mag at [3.6] and 19.1\,mag at [4.5]. This gives us a [3.6] depth of up to $-$8.5\,mag 
at 5\,Mpc and up to $-$11.5\,mag at 20\,Mpc.

\subsection{Follow-Up Ground-based observations}
We are undertaking concomitant ground-based surveys to monitor the SPIRITS galaxy sample 
in the near-IR and the optical at roughly a monthly cadence.  At the University of Minnesota's Mt Lemmon Observing Facility (MLOF), 
we use the three-channel Two Micron All Sky Survey  cameras \citep{Milligan1996,Skrutskie2006}  mounted on the 1.52m IR telescope \citep{Low2007}.
% to obtain a median depth of $J<$17.5\,mag, $H<$17.0\,mag and $Ks<$16.6\,mag.
At Las Campanas, we undertake near-IR
monitoring with the Retrocam on Dupont 100-inch telescope and optical monitoring using the
CCD on the Swope 40-inch telescope.  
At Palomar, we use the Samuel Oschin 48-inch (primarily $r$-band)
and Palomar 60-inch telescopes ($gri$-bands) for optical monitoring. Using the LCOGT network,
we obtain additional optical monitoring in $gi$-bands. In addition, follow-up of discovered transients 
was undertaken by a myriad of facilities including Keck, Magellan, Palomar 200-inch, SALT and RATIR.
%%REFS for each telescope  

\subsection{Follow-Up with the Hubble Space Telescope}
Following non-detections from the ground, we were able to set even deeper magnitude limits 
for two transients based on a small \HST\/ Director's Discretionary program (GO/DD-13935, PI H.~Bond).
We imaged SPIRITS\,14aje (in M101) and SPIRITS\,14axa (in M81) with the Wide Field Camera~3
(WFC3) in 2014 September. In the WFC3 UVIS channel, we employed an ``$I$''
filter (F814W), and in the IR channel we used ``$J$'' (F110W) and ``$H$''
(F160W) bandpasses. The sites of both targets had been observed with \HST\/ before their outbursts,
making it possible for us to compare our new images with the archival data. We
registered the {\it Spitzer} frames with the \HST\/ frames by measuring the
positions of isolated stars detected in both images, allowing the sites of the
transients to be located in the \HST\/ images to precisions of typically
$0\farcs1$ (More details in Bond et al. in prep).

\section{Software Pipelines}
\label{sec:software}

\subsection{Image Subtraction and Source Catalogs}

\begin{figure*}[!hbt]
\centering
\includegraphics[width=\textwidth]{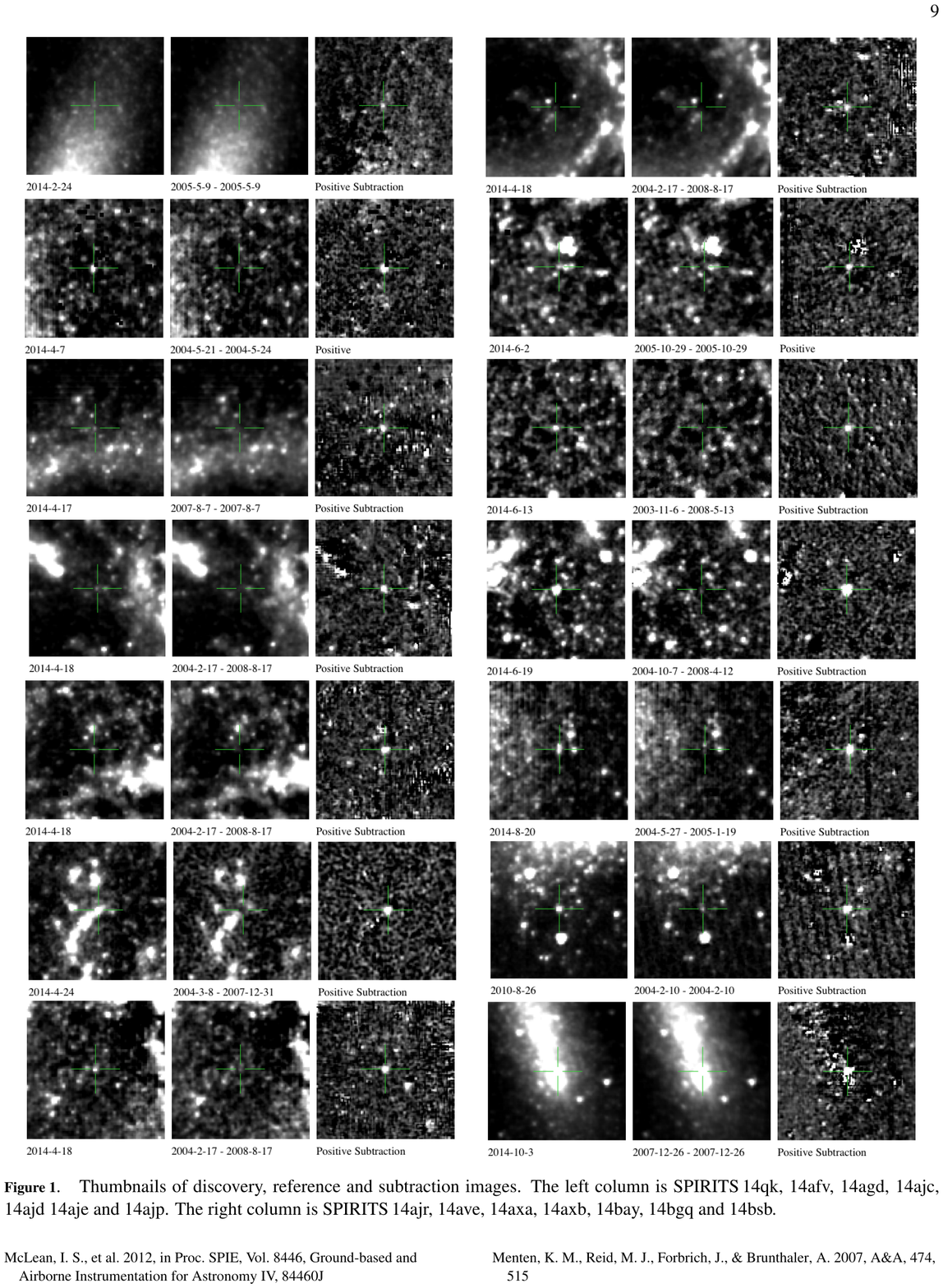}
\caption{ Thumbnails of discovery, reference and subtraction images. The left column is SPIRITS\,14qk, 14afv, 14agd, 14ajc, 14ajd 14aje and 14ajp. The right column is SPIRITS\,14ajr, 14ave, 14axa, 14axb, 14bay, 14bgq and 14bsb.
\label{fig:discovery}}
\end{figure*}

We construct reference images using archival {\it Spitzer} imaging using
supermosaics  or S4G stacks (\citealt{Sheth2010}; where supermosaics were unavailable) or
stacking prior ``bcd" observations in the archive  (where neither supermosaics nor S4G  stacks were available).
%If none were available, we used the first epoch of SPIRITS imaging as our reference. 
We use the ``maic" products of the {\it Spitzer} IRAC pipeline as our starting point for difference imaging. 
We adapted an image differencing and transient-source extraction code developed for the Palomar Transient Factory (PTFIDE\footnote{http://web.ipac.caltech.edu/staff/fmasci/home/miscscience/ptfide-v4.0.pdf}) to {\it Spitzer} imaging. The changes made to this software were (i) ability to operate on co-adds of individual IRAC exposures; (ii) masking of co-add--image regions with depths $<$5 exposures to mitigate cosmic rays and detector glitches; (iii) execution of the SExtractor tool (Bertin and Arnouts, 1996) to extract transient candidates from the difference images (as opposed to PSF-fitting for PTF); (iv) omission of dynamic photometric-gain matching between reference and science-image co-adds, and (v) a more streamlined and simpler PSF-matching scheme between images. Updates (iv) and (v) take advantage of the stable thermal environment of the {\it Spitzer} telescope. Examples of discovery image triplets are shown in Figure~\ref{fig:discovery}.

One may expect difference-imaging from space to yield fewer false positives than from a ground observatory where atmospheric conditions continuously modify the PSF between observations, however, we found this generally not to be the case with {\it Spitzer} (in part, due to the sparsely sampled PSF). The {\it Spitzer} difference-imaging is prone to a large number of false positives when the {\it Spitzer} field-of-view had a large rotation between the reference and science image epochs. The {\it Spitzer}-IRAC PSF profiles follow the fixed detector/optical diffraction patterns and these could not be easily matched and subtracted between rotated apparitions of the same piece of sky. Each of our candidates are visually vetted and we require at least two detections (in two filters or at two epochs) to weed out false positives. But, we caution that our search is incomplete and any inferred rates of transients are likely lower limits.  
 
\subsection{Forced Photometry}
We define a source to be transient if there is no detected quiescent point source underneath the source location in the reference frame (else, the source is a variable). To obtain magnitudes for transient sources, we perform forced aperture photometry on the subtracted images, assuming
zero flux present in the reference image. We sum the flux in an aperture with radius 4-pixels (2\farcs4) centered at the RA and Dec coordinates
determined by our transient detection routines. Sky background is measured within an annulus from 8 to 16 pixels
surrounding each source and subtracted from the total flux. Finally, fluxes are converted to magnitudes using the Warm {\it Spitzer}/IRAC
zero points of 18.8024 (channel 1) and 18.3174 (channel 2), along with aperture corrections of 1.21 and 1.22, respectively, as specified
by the IRAC instrument handbook. Since the subtraction images are noisy, we conservatively set the detection and upper limit threshold at 9$\sigma$. 

\subsection{Database and Dynamic Web Portal}
We architected a postgresql database to ingest the difference imaging products.  The search for transients is  
undertaken via a dynamic web portal. In 2014, {\it Spitzer} data was released every two weeks (the time-lag has now been reduced
to only a few days). Once the data is released, the SIDE pipeline is promptly run. 
Team members are assigned galaxies to look through candidate metadata and postage stamps to visually vet and flag interesting transients, typically within one day of data release. These transients are then assigned a name in sequential order by the database. For example, in the first year, 
a total of 131722 candidates (on 4396 new and archival subtraction images) were automatically loaded into our subtraction database. Of these, only 1693 
sources were assigned names and flagged for further inspection as transients or variable stars. Additional context information from various 
ground-based and space-based facilities is summarized on a source-specific webpage for reference. Interesting transients are announced via Astronomers
Telegrams (e.g., \citealt{ATEL1,ATEL2,ATEL3,ATEL4}).

\section{First Transient Discoveries}
\label{sec:discoveries}

In the first year, SPIRITS detected over 1958 variable stars and 43 IR transient sources. Of these 43 transients, 
21 were known supernovae and 4 were in the luminosity range of classical novae. SPIRITS supernovae have been discussed 
in-depth as a Type Ia sample \citep{Johansson2014}, a core-collapse sample \citep{Tinyanont2016}, and a case-study of the peculiar 
low-velocity SN\,2014dt \citep{Fox2016}.  Four transients had optical counterparts: SPIRITS\,14axm in NGC\,2403 (luminous blue variable;
van Dyk et al. in prep), SPIRITS\,14pz in NGC\,4490 (stellar merger candidate; \citealt{Smith2016}), SPIRITS\,14bme in NGC\,300 (high mass
X-ray binary; \citealt{Lau2016}) and SPIRITS\,14bmc in NGC\,300 (Adams et al. in prep). Here, we discuss the remaining 14 events (see Table~\ref{tab:radec}), which are 
unusual IR transients in the luminosity gap between novae and supernovae with no optical counterparts, hereafter referred to as SPRITEs 
(eSPecially Red Intermediate-luminosity Transient Events). 
   
\begin{figure*}[!hbt]
\centering
\includegraphics[width=\textwidth]{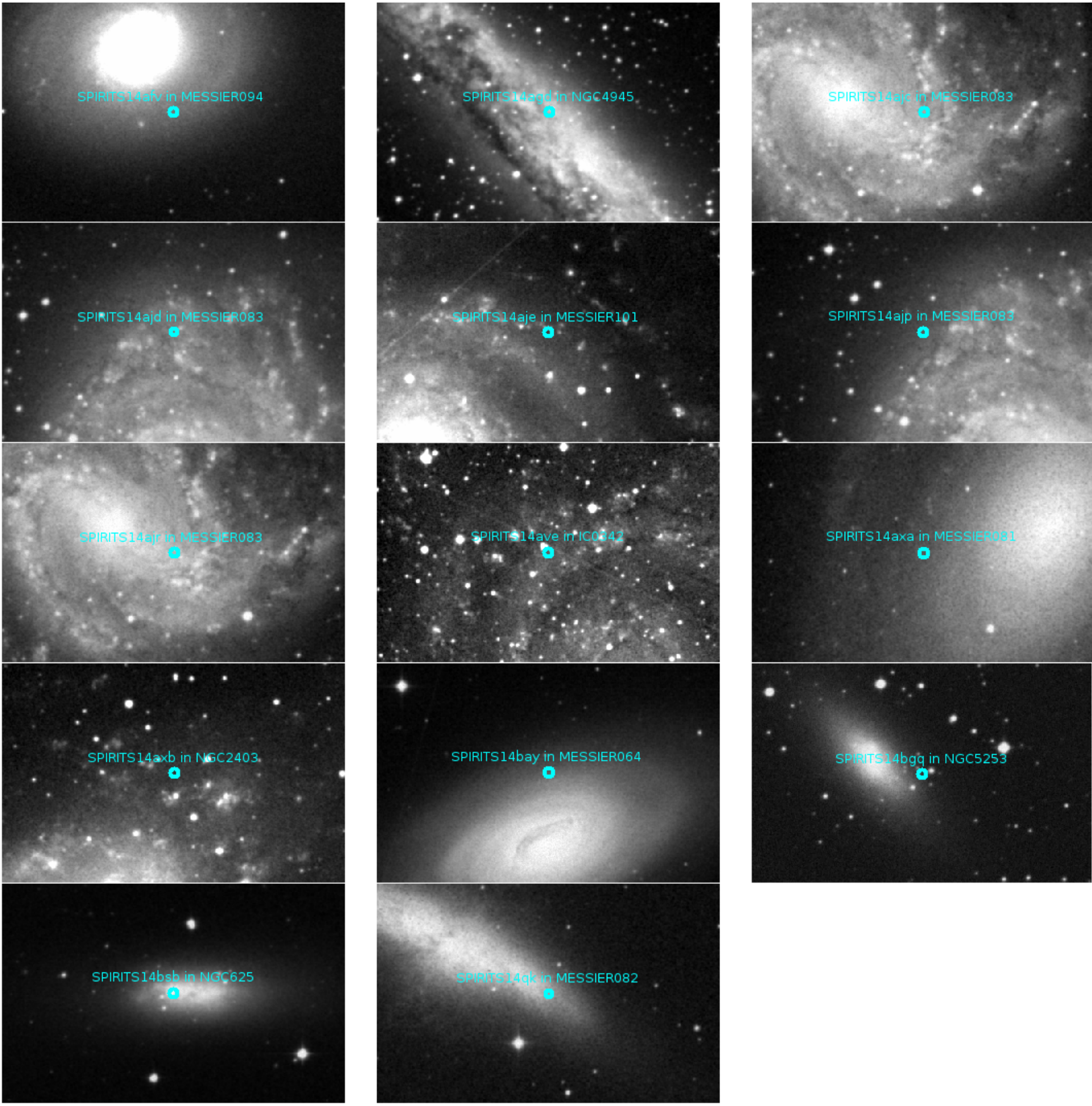}
\caption{Collage of host galaxies of 14 IR transients discovered by SPIRITS. Background images are 6.9' $\times$ 4.6', centered on the SPRITE location, and taken from the Digitized Sky Survey.  
\label{fig:hosts}}
\end{figure*}

The common properties of this new class of SPRITEs are: 
\begin{enumerate}
\item Peak luminosity at [4.5] brighter than $-$11\,mag and fainter than $-$14\,mag (see Table~\ref{tab:lcprop}).
%These limits are chosen to exclude classical novae and most supernovae. 
\item IRAC Color $[3.6]$-$[4.5]$ between 0.3\,mag and 1.6\,mag (see Table~\ref{tab:lcprop}).
%We note that foreground M-dwarfs and L-dwarfs are hotter than a color of 0.5\,mag \citep{Leggett2010} and T-dwarfs variability amplitude is limited to $<$0.2\,mag \citep{Heinze2013}.
\item No optical counterpart in concomitant or follow-up imaging to at least r$<$20\,mag (see Table~\ref{tab:followup}). 
%(iv) No detection in archival {\it Spitzer} reference imaging (progenitor limits in all bands)
\item Occurring in star-forming galaxies even though the SPIRITS galaxy sample has a mix of all galaxy types (see Figure~\ref{fig:hosts}) 
\end{enumerate}

\begin{figure*}[!hbt]
\centering
\includegraphics[height=0.57\textwidth]{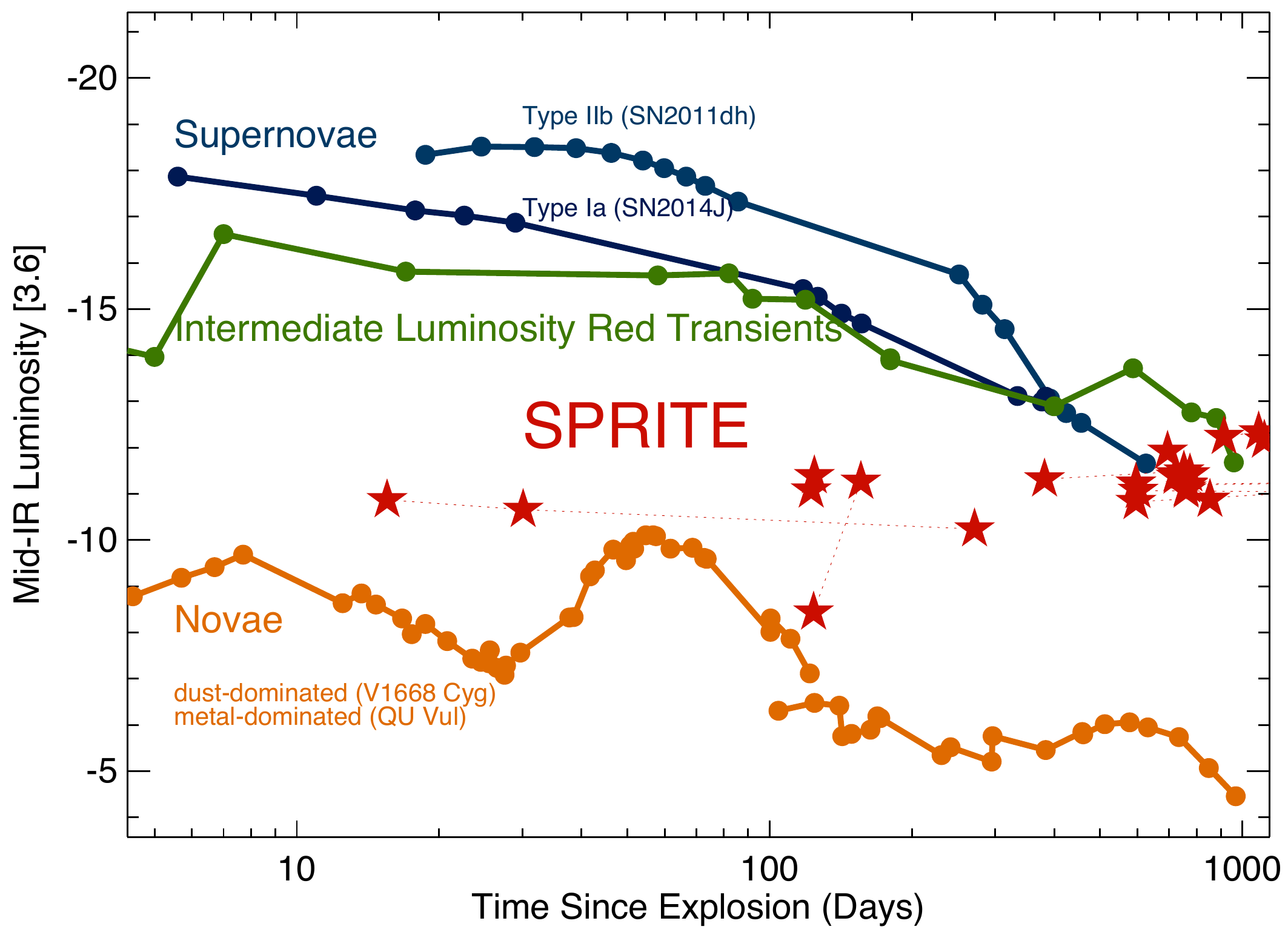} \\
\includegraphics[height=0.57\textwidth]{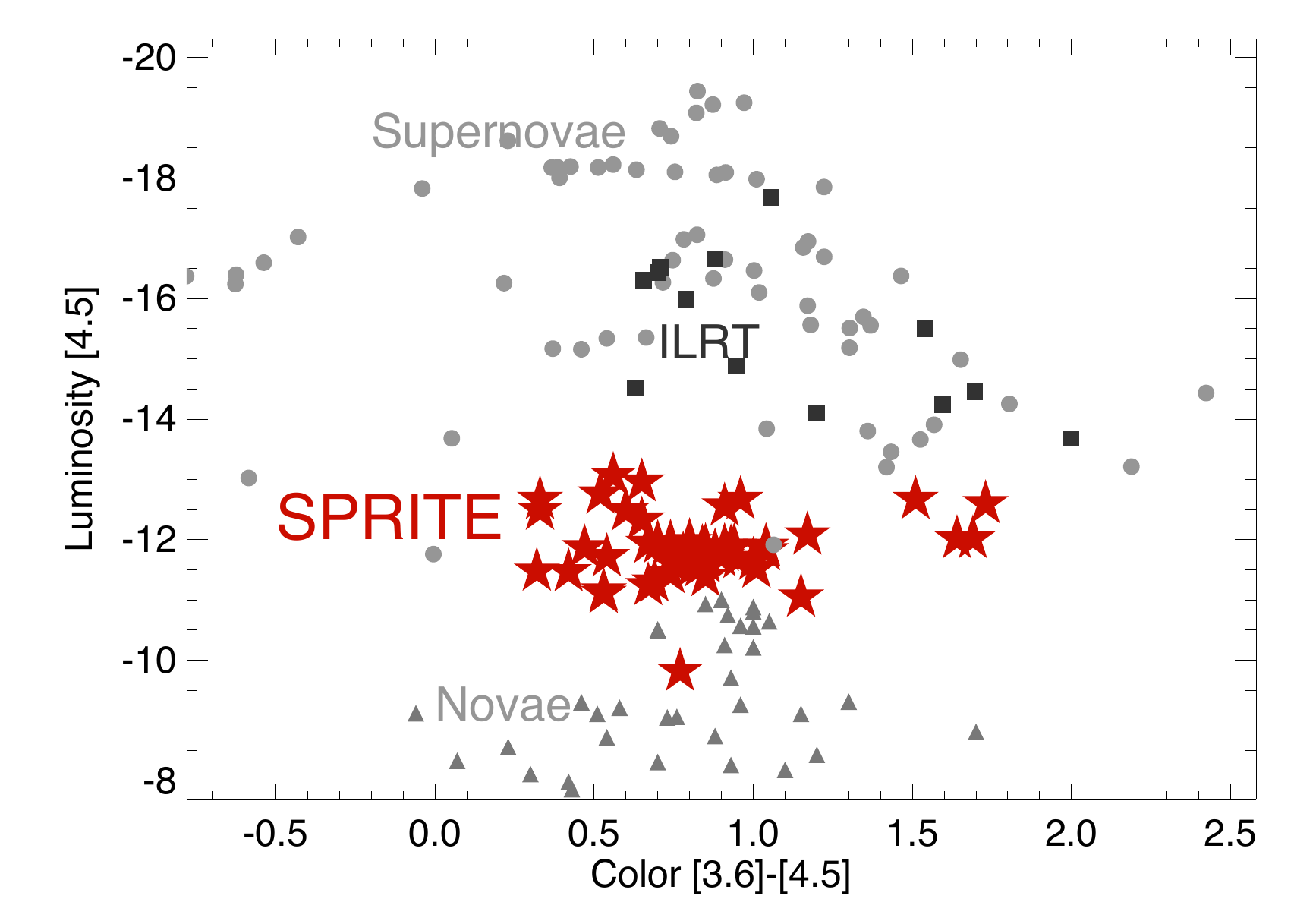}
\caption{ {\it Top: } Light curves of SPRITEs (red stars) are in the mid-IR luminosity gap between novae (orange) and supernovae (blue). Note that the assumed explosion time for SPRITEs is the last non-detection in archival data (and hence, the phase is a conservative upper limit).  {\it Bottom:} A luminosity-color rendition of the phase space of IR explosions illustrating the unique location of SPRITEs (corresponding effective temperatures are between 350\,K and 1000\,K\null). The comparison set includes all detections of all known supernovae hosted by SPIRITS galaxies \citep{Tinyanont2016,Johansson2014} and two Galactic classical novae \citep{Gehrz1995,Gehrz1980}. Detections at all epochs in the light curve for each transient is shown. 
%Blackbody light curves of SPIRITS transients indicates effective temperatures between 350\,K and 1000\,K.  Upper limits (downward triangles) are obtained from multiple telescopes including Keck, Palomar, Magellan, Dupont, Swope, MLOF, HST. 
\label{fig:lumcolor}}
\end{figure*}

First, we place SPRITEs in context of the IR transient phase space. We plot the IR luminosity evolution of SPRITEs 
and compare to well-known novae and supernovae (see top panel of Figure~\ref{fig:lumcolor}). SPRITEs are too luminous to be classical
novae.  The highest absolute magnitude that can be achieved by dusty classical novae is approximately $-$11\,mag at [3.6]. This maximum occurs when
an optically thick dust shell forms about 50--100\,days after outburst while the central engine is still near Eddington luminosity (e.g. NQ Vul, LW
Ser; see \citealt{Gehrz1995}).  At peak dust production, these novae can have a [3.6]-[4.5] color of $\approx$1.5\,mag (see \citealt{Ney1978,Gehrz1980}). 
Some very fast novae that do not form dust (e.g. V1500 Cyg; see \citealt{Gallagher1976, Ennis1977}) may also reach luminosities
of up to $\approx-$11\,mag for a few days at outburst.

%The upper mid-IR limit for dusty novae occurs when an optically thick
%dust shell that entirely covers the central engine forms about 50Ð100 days after outburst while the central
%engine is still near Eddington luminosity. In this case, the dust shell re-radiates the outburst luminosity.
%The highest absolute bolometric magnitude that can be achieved by thus far observed in a dusty nova
%dusty was ~ -7.3 (NQ Vul; Ney &amp; Hatfield 1978). When this nova formed an optically thick dust shell, its
%absolute [3.6] magnitude, M 3.6 , was ~ -11.3. Other similar examples cited in Gehrz et al. (1885) were LW
%Ser (M Bol ~ 6.75; M 3.6 ~ -10.35) and V705 Cas (M Bol ~ 7.1; M 3.6 ~ -10.4).
%classical novae is $-$11 mag at [3.6]. 

\begin{figure*}[!hbt] 
\centering
\includegraphics[height=\textwidth,width=\textwidth]{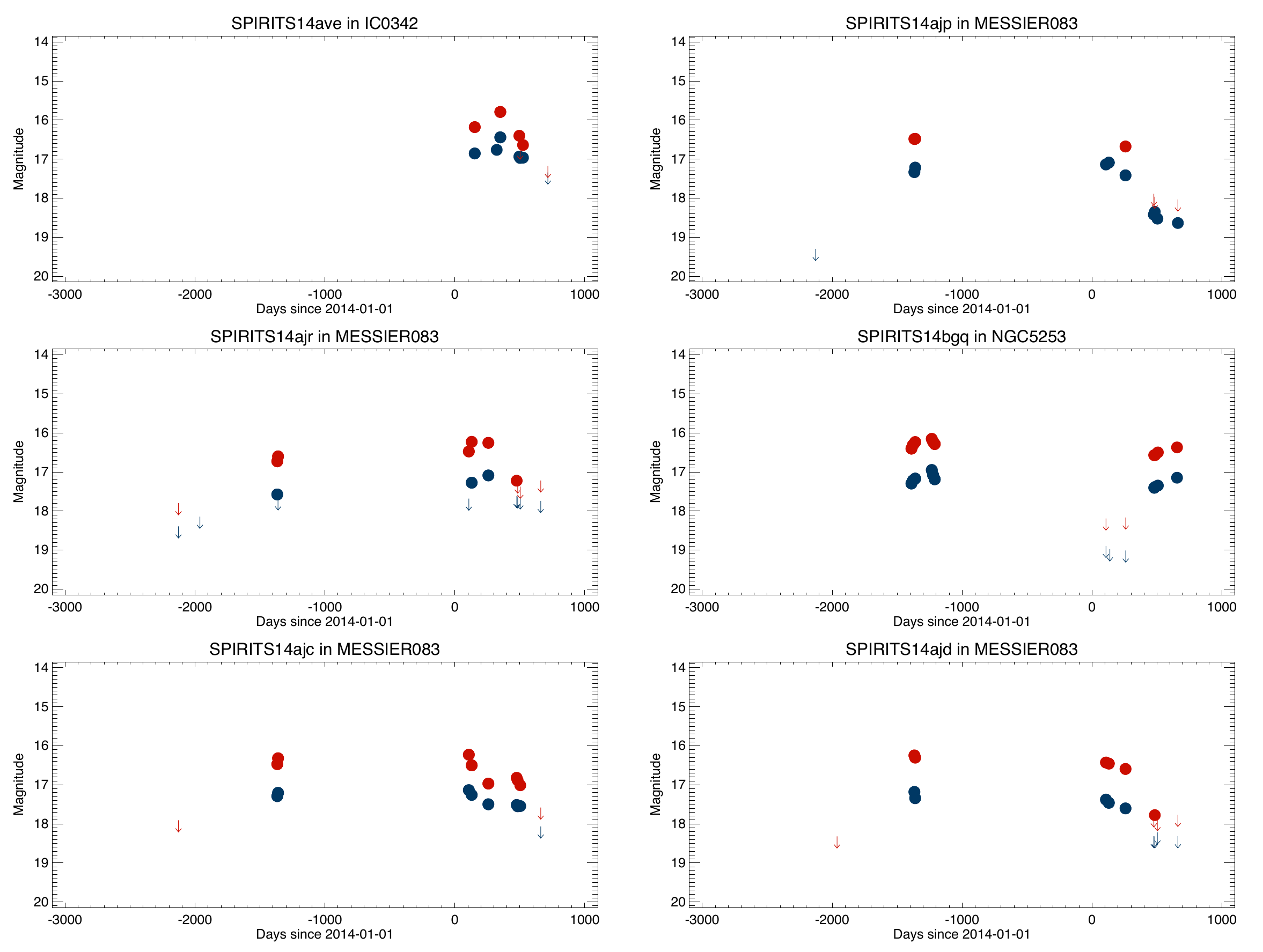}
\caption{IR light curves of relatively slow evolving SPRITEs.}
\label{fig:fastlc}
\end{figure*}

Next, we characterize the light curves of SPRITEs. The {\it Spitzer} light curve data are presented in Table~\ref{tab:allphot}. 
We broadly categorize SPRITEs into two relative speed classes (see Table~\ref{tab:lcprop}): 
(i) six SPIRITEs evolve slowly over many year timescales with speeds slower than 0.5 mag yr$^{-1}$ (Figure~\ref{fig:slowlc}), 
(ii) eight SPRITEs evolve faster than 0.5 mag yr$^{-1}$ (Figure~\ref{fig:fastlc}). Of these, four evolve on few month timescales, 
faster than 1.5 mag yr$^{-1}$ and fade below detectability in less than one year. 

But, the light curve comparison is limited as the light curves of SPRITEs are diverse, infrequently sampled and the explosion times 
are poorly constrained. Therefore, we plot IR luminosity versus IR color at all epochs (bottom panel of Figure~\ref{fig:lumcolor}).  We find that 
SPRITEs occupy a unique region in phase space between novae and supernovae on this dynamic HR diagram of explosive IR transients. 
We note that the IR colors of SPRITEs are as red as the reddest novae, core-collapse supernovae and ILRTs. The corresponding effective 
black body temperatures of SPRITEs span 350\,K to 1000\,K. 

The puzzling absence of optical emission from SPRITEs challenges a supernova interpretation.
Typical supernovae are brighter than 20\,mag at 20\,Mpc for many months in the visual wavebands. Thus, no prior
history of optical detection (despite the intensive monitoring of nearby galaxies by several synoptic surveys and amateurs 
in the optical wavebands) suggests that SPRITEs are unlikely to be old supernovae. 
If the IR explosion time is strongly constrained and the lifetime is shorter than a year, it implies that the peak luminosity was not much higher 
than observed (and, that the transient is not old). We consider the hypothesis that two fast SPRITEs --- 
SPIRITS\,14axa and SPIRITS\,14bay ---  are obscured supernovae.  Applying a simple extinction law to a 
10,000\,K blackbody \citep{Cardelli1989, Chapman2009}, we find that an observed [3.6]-[4.5] color of 1\,mag requires 
visual extinction of 30\,mag respectively! This is difficult especially given the location of SPRITEs in the middle-to-outer 
galaxy disks (Figure~\ref{fig:hosts}). Moreover, if this is correct, given that there were six new optical supernovae in the SPIRITS 
sample in 2014, it would suggest that optical surveys are missing a fourth of the supernovae due to obscuration. 

%Furthermore, inconsistent with our observations, this would imply the supernova is brighter than $-$15\,mag to 
%$-$12\,mag even at 3.6um. [CHECK NUMBERS] 
%http://www.pas.rochester.edu/~emamajek/memo_ir_reddening.html
%http://casa.colorado.edu/~ginsbura/filtersets.htm

%(ii) Detection in both IRAC filters (to eliminate contamination by moving objects in the solar system) 
%(iii) No previously known optical transient, radio or X-ray counterpart at this position (to eliminate known supernovae, AGN)
%(iv) No optical counterpart in concomitant or follow-up imaging. 
%(v) IRAC Color $>$0.5\,mag to minimize foreground dwarf contamination. We note that foreground M-dwarfs
%and L-dwarfs are hotter than 0.5\,mag (Legget et al. 2010) and T-dwarfs variability is limited to $<$0.2\,mag (REF). 
%(vi) Amplitude $>$0.5\,mag relative to archival {\it Spitzer} reference imaging

Next, there are a few possible theoretical models that may explain SPRITEs. For example, coalescence of 1--30 M$_{\odot}$ 
binaries is expected to create copious amounts of dust in an optically thick wind launched during the stellar merger \citep{SokerTylenda2006,Ivanova2013,Nicholls2013,Pejcha2016}.  
Another possibility is that these are electron-capture induced collapse of 8--10 M$_{\odot}$
extreme AGB stars where the shock breakout did not destroy all the dust surrounding the progenitor \citep{Kochanek2011}. 
Yet another possibility is that weak shocks in failed supernovae that form black holes may also not lead to bright optical transients but
rather IR transients as the ejection of large amounts of material at low velocity may
condense to form dust \citep{Piro2013,Lovegrove2013}. We consider the likelihood of each of these models below. 

The speed of evolution is diagnostic of the origin in that a terminal, explosive event (eCSNe, obscured SNe,  failed SNe) would likely belong
to the fast class and a stellar merger would likely belong to the slow class.  Stellar merger of massive stars may 
result in a slowly evolving red remnant. As was seen with SPIRITS\,14pz in NGC\,4490 \citep{Smith2016} and the transient in M\,101 \citep{Blagorodnova2016}, 
the SPRITE luminosities are consistent with late-time observations of a stellar merger. The young stellar population in
the vicinity of these transients also supports the presence of massive stars. Five slow SPRITEs could be stellar mergers 
(We exclude SPIRITS\,14bgq as it repeats -- appears, disappears and re-appears -- and could either be a background AGN or an extreme AGB variable).
If this hypothesis is correct, the rates appear to be higher than that estimated by \citealt{Kochanek2014}. 

Now for the relatively fast events, the nature of the explosion can be disentangled if there is a progenitor identification.
Distinguishing features of the eCSNe class are: (i) the detection of an IR progenitor
star with an absolute magnitude brighter than $-$10\,mag and color redder than 1\,mag \citep{Thompson2009, Kochanek2011}, and (ii) subsequently,
a monotonic decline of the transient emission below the progenitor luminosity \citep{Adams2016}.
Only two SPRITEs have detections in archival {\it Spitzer} imaging: SPIRITS\,14bgq and SPIRITS\,14bsb. But neither is a promising
eCSN as SPIRITS\,14bgq is eruptive (not explosive) and SPIRITS\,14bsb appears to be part of a massive star cluster.
The remaining SPRITEs have upper limits deep enough (see Table~\ref{tab:lcprop}) to rule out most of the progenitor parameter 
space delineated by \citealt{Thompson2009}.

If an HST progenitor is detected and consistent with a massive star, there are two possibilities: an eruption event like eta Carinae
or a terminal explosive event like the formation of a stellar mass black hole. Both would be consistent with a young population. One way to distinguish 
these two would be to continue monitoring to see if the fading is monotonic (hence, terminal) or if there is another episode of eruption.
But none of the SPRITEs with both pre-explosion and post-explosion archival HST imaging have candidate progenitor counterparts.  
Specifically, our HST imaging of SPIRITS\,14aje in 2014 September shows only one faint star detected at the edge of a 3$\sigma$
positional error circle in $I$-band. But this star does not vary between 2003 and 2014, is not detected in the $J$-band and $H$-band
and likely unrelated to the transient. Similarly, our HST imaging in $I$-band of SPIRITS\,14axa shows one star that brightened by $\sim$0.4~mag 
between 2002 and 2014, but it is consistent with a normal field red giant lying close to the tip of the red-giant branch and likely unrelated.

In summary, the IR photometric data alone is not sufficient to  distinguish between 
various models. Ground-based spectroscopic data has been difficult to obtain because these transients
are either very faint or not detected in the near-IR/optical wavelengths. 
Next, we present a detailed case-study on one transient for which we see shock-excited emission lines 
in a near-IR spectrum and hence, infer a physical origin. 

\section{SPIRITS\,14ajc: A transient driving a shock into a molecular cloud}
\label{sec:14ajc}

SPIRITS\,14ajc  in M\,83 went into outburst in 2010 and has 
stayed at a $[3.6]$-band luminosity of  $-11$~mag and a [3.6]$-$[4.5] 
color of 0.7~mag for the past four years (Figure~\ref{fig:slowlc}). 
No quiescent source was detected in {\it Spitzer} images taken between 
2006 and 2008.    No optical or near-IR counterpart is detected 
in ground-based follow-up in 2014 (see Table~\ref{tab:followup}).  
%Moreover, {\it HST\/} images taken in 2012 (during the outburst), 
%with the Wide Field Camera~3 in F814W (``$I$'' band) and F160W 
%(``$H$'' band) filters, failed to detect a source brighter than $I>$25.5~mag 
%and $H>$22~mag. 
Based on the cool SED of the transient, the effective 
blackbody temperature is approximately 900\,K. 

SPIRITS\,14ajc is located in a spiral arm of M\,83 with intense star 
formation (Figure~\ref{fig:hosts}).    The  2.6 mm \Tco\  J=1$-$0
emission at the position of SPIRITS\,14ajc, 
measured  with the Nobeyama Millimeter Array \citep{Hirota2014}, 
has a  peak brightness temperature of  2.8 K,  a velocity integrated
flux $I(CO)$ = 28.6 Kelvin~km~s$^{-1}$ in the
6\arcsec\ by 12\arcsec\  synthesized beam, and is  centered at
\Vlsr\ = 572~$\pm$~15 \kms .  The X-factor method 
\citep{Bolatto2013} to convert  $I(CO)$ into H$_2$ column density  
using a Solar-vicinity value for the X-factor ($2.0 \times 10^{20}$ 
cm$^{-2}$~K$^{-1}$) gives  N(H$_{2}$) $\approx$ 5.7$\times$10$^{21}$ cm$^{-2}$ %[3.6]
per beam  corresponding to a visual-wavelength (V-band)  extinction of 
6\,mag if the cloud were distributed  uniformly over the beam.
The NMA beam-size corresponds to a physical scale of 128\,pc $\times$ 279\,pc 
at the distance of M\,83, making it likely that the extinction
is patchy and potentially higher along the line of sight to SPIRITS\,14ajc.  

A  $K$-band spectrum obtained with the MOSFIRE spectrometer \citep{McLean2012} on the Keck I 
10-meter telescope on 8 June 2014 found  five  emission lines of 
molecular hydrogen, and neither any continuum nor any other lines.     
The observed properties of these ro-vibrational transitions 
are summarized in Table~\ref{tab:14ajc}.    The velocities are consistent with the CO  
radial velocity at the position of SPIRITS\,14ajc.    The relative line intensities  of the four v=1$-$0 
ro-vibrational transitions are consistent with shock-excitation at a temperature 
around 1000 to 2,000 K  and similar to the relative intensities in Herbig-Haro 
objects such as HH~211 \citep{OConnell2005}.  
However, the intensity ratio of the v=2$-$1 to 1$-$0  
vibrational  transition corresponds to a higher temperature, possibly indicating that
fluorescent pumping in the ultraviolet Lyman and Werner bands  
may play a role in exciting the higher vibrational states of H$_2$, or
 that the shock structure in the emission regions has a 
more complex temperature structure.    

\begin{figure*}
\begin{center}
\includegraphics[width=0.8\textwidth]{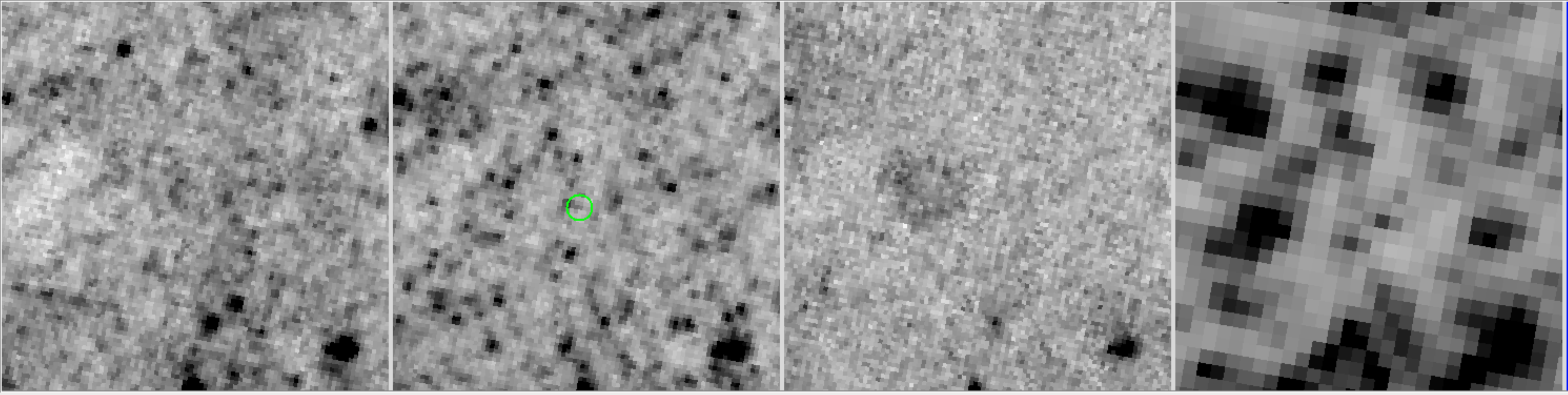}
\includegraphics[width=0.8\textwidth]{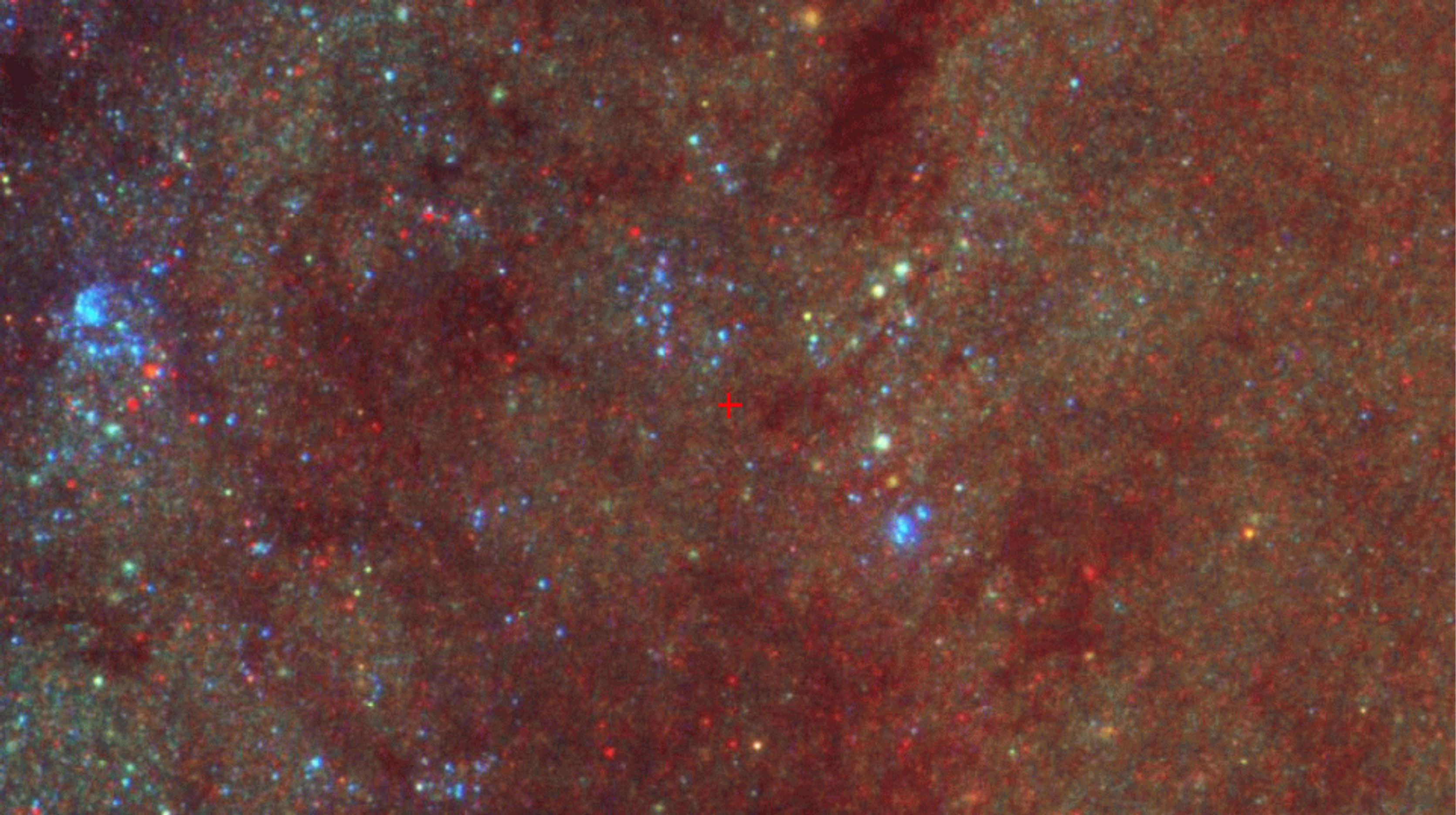}
\figcaption{
{\it Top:} {\it HST\/} WFC3 images of the site of SPIRITS\,14ajc in M83, taken in 2012 while
the event was underway. Frames are $3\farcs6$ high, and have north at the top
and east on the left. From left to right the frames are in $V$ (F547M), $I$
(F814W), H$\alpha$+[\ion{N}{2}] (F657N), and $H$ (F160W).  The green circle in
the F814W image marks the position of the transient, with a radius corresponding
to the 3$\sigma$ positional uncertainty (alignment RMS was $0\farcs04$). There is a very faint star near the
center of the error circle in the F814W image, and a brighter one at the eastern
edge, but both are unlikely to be the counterpart of 14ajc. Neither
star is detected at $H$. A faint bubble-like emission nebula lies close to 14ajc
in the third frame, but outside the error circle.
{\it Bottom:} Color rendition showing the stellar and ISM environment around SPIRITS\,14ajc in
M83, made from \HST\/ images in $U$, $B$, and $I$ in the Hubble Legacy Archive.
The field is $20''$ high, with north at the top and east on the left. The
location of SPIRITS\,14ajc is marked with a red cross. The transient lies near several
young associations and dark dust lanes, but apparently not within them.
}
\label{fig:hst14ajc}
\end{center}
\end{figure*}

The site of SPIRITS\,14ajc in M83 was fortuitously imaged by \HST\/ with the
WFC3 camera in 2012 (program GO-12513, PI W.~Blair). 
%For detailed examination,
%we chose frames taken in an intermediate-band $V$ filter (F547M), $I$ (F814W),
%narrow-band H$\alpha$+[\ion{N}{2}] (F657N), and $H$ (F160W). The F814W and F657N
%images were obtained on 2012 July~22, and the other two on 2012 September~3.
%These dates are 528 and 485 days before 2014 Jan~1, respectively; 
The SPIRITS\,14ajc light curve suggests that the event was underway in 2012, 
although unfortunately there were no {\it Spitzer} observations in this year.
%We aligned a \Spitzer\/ image taken on 2014 April~18, when 14ajc was brightest,
%with the WFC3 F814W frame in order to locate the site of the transient. The
%registration was done using seven isolated stars detected in both frames,
%yielding a positional scatter of $0\farcs04$.  
Figure~\ref{fig:hst14ajc} depicts the site of SPIRITS\,14ajc in the \HST\/ $V$, $I$, H$\alpha$+[\ion{N}{2}], and $H$ bandpasses. 
%In the $I$ frame, the green circle has a radius corresponding to the 3$\sigma$ position uncertainty.
The source lies in a very crowded stellar field with only modest extinction. In
the $I$-filter, there is a faint star (26.0\,mag, $M_I\simeq-3.8$) which lies close to the center of the
error circle and a brighter star (25.0\,mag, $M_I\simeq-2.8$) which lies at the eastern edge of the 3$\sigma$ position
uncertainty. The absolute magnitudes are consistent with both stars being normal field red
giants. Only the brighter of these two stars is marginally detected at $V$and neither star is detected in $H$-band. 
In the absence of any other \HST\/ images in these filters taken at different dates, we cannot definitely rule out that either of
the $I$-band sources is the counterpart of SPIRITS\,14ajc, but the lack of a detection at
$H$ argues against this. The 5$\sigma$ limiting magnitude (Vega-scale) for the
$H$-band exposure is about 23.5. The H$\alpha$+[\ion{N}{2}] image (third frame)
shows nothing at the site, but nearby is a small, faint, bubble-like
emission nebula just outside the error circle. Its diameter is about $0\farcs6$,
or approximately 16~pc. To characterize the stellar and ISM environment, 
we show a color rendition of its wider surroundings.
%, derived from the HLA using images from GO-12513 obtained in $U$, $B$, and $V$. 
There are several young associations, some containing red supergiants, as well as a network of dark dust
lanes, within a few arcseconds of the site. However, 14ajc is not clearly
associated with either the very young clusters or dark dust features.

We consider two models for SPIRITS\,14ajc:   the explosion of a supernova
immediately behind or inside a dense molecular cloud \citep{Kasliwal2005}, 
and the production of an eruptive protostellar outflow similar  to
the explosion that occurred in Orion about 500 years ago 
\citep{Zapata2009,Bally2011,BallyGinsburg2015}. 

\subsection{Is SPIRITS\,14ajc powered by a supernova?}

In the supernova scenario,  two mechanisms may contribute to the IR signal.
First, the flash produced by the explosion can be reprocessed into
the IR portion of the spectrum by dust as the light echo propagates 
through the dense interstellar medium.  In this scenario, foreground dust  completely
extinguishes the visual to near-IR wavelength signature of the event.   
%The \hh\ emission in this picture is produced by fluorescent excitation by UV light from the supernova.   
Second, if the supernova is sufficiently close to a dense cloud, the impact of the forward shock can
excite  \hh\ emission as the  blast-wave slams into a dense molecular medium.   
There may be two components to this emission: collisional excitation of \hh\  in the 
swept-up, compressed, and accelerated, post-shock layer, and fluorescent excitation
by  the UV radiation emitted by fast shocks.    Shock radiation can excite both  
\hh\ located ahead of the shock,  and surviving or re-formed \hh\ in the swept-up,
post shock layer.   A  visual  extinction of  more than 15\,mag is required to hide the visual
and near-IR continuum of a supernova.    

The ionizing radiation of the supernova progenitor  
should have created a large ionized cavity in the surrounding ISM.  
However,  there is no obvious HII region at the location of SPIRITS\,14ajc.    
It is possible that foreground dust obscures any cavity or HII region.   
Alternatively,  a high-velocity,  ` runaway'  OB-star would have only carved a small cavity.   
More than 30\% of massive  OB stars are ejected  from their birth sites with velocities 
greater than 20 \kms, and more than 10\% with  velocities larger  than 100 \kms\  
\citep{GiesBolton1986}.   Two ejection mechanisms have been identified:  
dynamical  interactions such as the re-arrangement of the 
non-hierachical multiple stars into a hierarchical configuration such as a compact binary 
and ejected members.  \citep{Hoogerwerf2000,Hoogerwerf2001,Gualandris2004}, and 
the  supernova explosion of the most massive member of an OB-star binary which
results in the ejection of the surviving member at its pre-supernova Keplerian orbital speed. 
Alternatively, if the runaway O star were in a red-supergiant phase prior to its demise as a supernova
as it entered a molecular cloud, it would have avoided the production of an ionized cavity. 
It is also possible that the supernova was a Type Ia which happened to drift into
a molecular cloud.  However, such events must be rare since the volume filling factor
of dense molecular clouds in a galaxy tends to be less than 1\%.

\subsection{Is SPIRITS\,14ajc powered by a massive protostellar eruption?}

In the second model, the SPIRITS\,14ajc  transient may trace an explosion triggered by
either a  protostellar collision or a violent  dynamical interaction of several massive
protostars \citep{ Davies2006,Dale2006}.     Such an event is 
suggested to have occurred in the OMC1 cloud core in the Orion A molecular cloud
located behind the Orion Nebula, the nearest region of ongoing massive star formation 
(D $\approx$ 414 pc; \citealt{Menten2007}).   The OMC1 outflow consists of a spectacular,  
wide opening-angle,  arcminute-scale (0.1 to 0.3 pc) outflow which is the brightest
source of near-IR H$2$ emission in the sky  \citep{AllenBurton93,Kaifu2000,
Zapata2009,Bally2011,BallyGinsburg2015}.   The radial velocity of the brightest part 
of the \hh\  emission exhibits a line-width of less than about 70 \kms , consistent with 
the \hh\ line-widths of SPIRITS\,14ajc.  Radial velocities and proper motions of more than 300
\kms\ are observed in visual and near-IR spectral lines such as [OI], [SII], and [FeII].  
The  momentum and kinetic  energy  is at least  
$160$ M$_{\odot}$ km s$^{-1}$ and  $4 \times 10^{46}$  ergs  \citep{Snell1984} to  
$4 \times 10^{47}$ ergs  \citep{KwanScoville1976}.     \citet{Zapata2009}  presented 
a CO J = 2--1 interferometric study and found a dynamic age of  about $500$ 
years for the larger OMC1 outflow.   The initial explosion energy required to drive
the observed outflow is between $10^{47}$ to  $10^{48}$ ergs.  

High-velocity, runaway stars are common among massive stars 
\citep{Hoogerwerf2000,Gualandris2004}.
Radio-frequency astrometry has shown that two radio-emitting stars in OMC1, 
the 10 to 15 \Msol\  Becklin-Neugebauer (BN) object, and radio source I, thought
to have a mass of about 20 \Msol\ \citep{Goddi2011},  have  proper motions  of 
25 and 14  km s$^{-1}$ respectively away from a region less than 500 AU in diameter 
from which they were ejected  about  500 years ago \citep{Rodriguez2005, 
Gomez2008,Goddi2011}.  The kinetic energy in stellar motions is about 
$4 \times 10^{47}$ ergs.   The total energy of the OMC1 event, $\sim 10^{48}$ ergs,
consists of the  kinetic energy in the ejected stars,  the current kinetic energy of the 
outflow, and  the energy radiated away by shocks over the last 500 years   
\citep{Bally2011}.
 
\citet{BallyZinnecker2005} proposed that the OMC1 explosion was triggered 
by the collision and merging of forming massive stars.  
Dynamic friction and Bondi-Hoyle accretion onto massive protostellar core inside 
a dense cluster-forming clump may have lead to rapid  migration of the most 
massive protostars to the center of the clump's potential well where they formed a
non-hierarchial system of massive stars with similar interstellar separations.    
Such systems are unstable;  interactions led to the formation of  a 
hierarchical system  consisting of a  compact binary and a distant  third member
$\sim500$ years ago.  
In this scenario,  the final 3-body encounter would have resulted in the 
formation  of a compact, AU-scale binary,   most likely source I, and the ejection 
of both radio source I and BN from the OMC1 core \citet{Bally2011} . 

\citealt{Goddi2011} used N-body simulations to show  that the most likely initial configuration
of massive stars in OMC1 that led to the observed current configuration of ejected stars
consists of a massive binary star interacting with another massive star.  
The interaction leads to a hardening of the binary and consequent release of gravitational 
potential energy. 

Binary stars are common among massive stars and the mass distribution is peaked 
at similar component masses.  A massive binary can form from the bar instability in a 
massive disk.   This mechanism tends to produce circular orbits in the plane of 
cirumbinary and/or  circumstellar  disks.    In forming clusters with a high density of 
protostars,  massive stars surrounded by massive disks also have a high probability 
of capturing clusters members to form binaries 
\citep{MoeckelBally2006, MoeckelBally2007a}.   The secondary in such a 
capture-formed binary is most likely to be on a high-eccentricity, high-inclination orbit  
with respect to the spin axis of the massive stars and its disk  
\citep{MoeckelBally2007b, Cunningham2009}.     Thus, binary-binary or binary-single star
interactions are likely in dense proto-clusters. 

The kinetic energy of the outflow and ejected  stars came from the release 
of gravitational binding energy of a compact binary  formed  by the  
interaction of  three or more stars.  The total  energy liberated by the formation or
hardening of a binary depends on the stellar masses, $M_1$ and $M_2$,  and the
orbit semi-major axis, $R$, as $E_B \approx G M_1 M_2 / 2 R$.
For stars in the mass range 10 to 100 \Msol\ and binary separation 0.5 $<R<$10 AU
$E_B$ ranges from  $10^{47}$ to $ 10^{51}$ ergs. 
Assuming that radio source I consists of a pair of 10 \Msol\ stars and that the energy 
required to eject the stars, the outflow, and to account for radiative losses is 
$E = 10^{48}$ ergs,   the semi-major axis of the final binary must be
$R \sim GM^2 / 2 E \approx$ 0.9 AU. 

Forming massive stars accreting at rates of $\sim 10^{-4}$ to $\sim 10^{-3}$ \Msol\  yr$^{-1}$
tend to have bloated, AU-scale photospheres and resemble red supergiants
\citep{HosokawaOmukai2009}.    Radio source I has a photospheric temperature of
$\sim 4000$ K, consistent with high-accretion models \citep{Testi2010}.    If at least
one star before the interaction were accreting at such a high rate,  the $10^{48}$ erg
energy requirement of the OMC1 event implies that any attempt to form an AU-scale 
binary would have led to a protostellar collision and consequent ejection of some of the
bloated star's photosphere before ejection from the cloud core.

In the dynamical interaction model, the disruption of circumstellar disks by the final close-in 
3-body stellar encounter, combined with the recoil of the larger-scale envelope 
power the OMC1 outflow and a luminous IR flare.    Ejected material  slams into 
the surrounding cloud core and lower-density envelope with speeds 
comparable to the Kepler speed at their point of origin.  Ejecta velocities 
are expected to range from  10s of \kms\ for material originating tens of AU to over 
500 \kms\ for ejecta from within a few tenths of an AU of a massive star.  The X-ray, UV, 
and visual light from shocks will be obscured and reprocessed by the surrounding  
cloud into the IR.   Powerful shocks can produce
an IR  flare with luminosities up to $10^{51}$ ergs for the most massive star 
collisions \citep{BallyZinnecker2005}. 

The duration of the IR-flare is given by the shock crossing  time of the part of 
the clump  sufficiently dense to reprocess the shock-energy into the IR.
The kinetic energy of the outflow is converted by surrounding dust
to the mid and far-IR where the clump density is sufficiently large. 
Taking a clump radius, $R_{\rm clump} \sim$ 0.01 $-$ 0.05 pc and an initial ejecta velocity of 
$V_{\rm ejecta} \sim$ 500 \kms\  implies a crossing time 
$t_{\rm cross} \sim R_{\rm clump} / V_{\rm ejecta}$
$\sim$ 20 to 100 years.   Emission of $2 \times 10^{47}$ ergs in 20 years, a lower bound 
on the early radiative losses for an Orion-like event, implies a mean
luminosity of $10^5$ \Lsol .

The Orion OMC1 explosion is not unique among star forming regions.  
\citet{Zapata2013} found evidence for a powerful explosive outflow in the DR 21 complex  
in Cygnus.    If such dynamic interactions are responsible for  the large number 
of runaway O stars, and binaries among massive stars, the event rate of OMC1-like 
events ought to be comparable to the birth-rate of massive stars.  We propose that
SPIRITS\,14ajc may trace the IR flare produced by a dynamic interaction of
forming massive stars that either led to the formation of or hardening of a compact, AU-scale
binary, or possibly a protostellar merger.

\section{Conclusion and A Way Forward}
\label{sec:conclusion}

SPIRITS has discovered a class of unusual IR transients called SPRITEs. SPRITEs are in the luminosity gap between 
novae and supernovae, with mid-IR [4.5] absolute magnitudes between $-$11 to $-$14\,mag (Vega) and [3.6]-[4.5] colors 
between 0.3\,mag and 1.6\,mag. The photometric evolution of SPRITEs is diverse, ranging from $<$0.1 mag yr$^{-1}$ to $>$7\,mag yr$^{-1}$. 
SPRITEs appear to represent diverse physical origins and each one merits an in-depth investigation to decipher its nature.
Next, we address challenges encountered in the first year of the SPIRITS survey and efforts to overcome them. 
%Specifically, SPIRITS\,14ajc  in Messier\,83 shows excited molecular hydrogen emission and
%%suggesting that it represents the birth of a massive star binary driving a shock through a molecular cloud.  
%may represent an explosion triggered by the dynamic decay of a non-hierarchical system of massive stars
%that led to either the formation of a binary or a proto-stellar merger. 

A challenge in deciphering the nature of SPRITEs was the sparse sampling of their mid-IR light curves
(due to the small number of epochs) with poor constraints on their explosion date (relative to archival images taken
many years ago). Subsequent transients discovered in later years of the SPIRITS survey have both a
better cadence on their light curve (1 week and 3 week cadence baselines were added) 
and better constraints on their explosion date (as the 2014 SPIRITS data serve as a reference). 

Another challenge in the first year was the difficulty of immediate ground-based follow-up due to the delay of availability of {\it Spitzer}
data and the mismatch between {\it Spitzer} and ground-based visibility. Both of these have improved in later years with 
Spitzer introducing early data release for time-critical programs and the additional SPIRITS
cadence baselines being scheduled preferentially in time windows that facilitate ground-based follow-up.  
 
Yet another challenge was identifying possible progenitor stars due to a combination of 
a dense stellar population in these nearby galaxies and the coarse {\it Spitzer} resolution. In order to address this, 
we now have an ongoing Hubble Space Telescope program to image these IR transients while they
are still active.

Our study of  SPIRITS\,14ajc highlights the importance of spectroscopy in solving the mystery of the physical nature
of the transient. Specifically, detecting the excited molecular hydrogen lines pointed to a shock driven by the dynamical
decay of a non-hierarchical system of massive stars. However, spectroscopy has also been extremely challenging as these transients are
too red for ground-based instruments, {\it Spitzer} is too warm now for mid-IR spectroscopy and SOFIA is not sensitive enough for these
faint transients. Spectroscopy with the James Webb Space Telescope (JWST) of slowly evolving SPIRITS transients 
would shed light on their nature. Specifically, the low resolution spectrometer could determine dust mass, 
grain chemistry, ice abundance and energetics to disentangle the proposed origins.
   
Regardless of the open questions on their origin, the discovery of SPRITEs representing a new class(es) of IR transients at a rate 
comparable (or higher) than supernovae is encouraging and motivates a synoptic IR search. However, undertaking a wide-field IR search for transients
beyond targeting nearby galaxies requires overcoming the formidable challenge posed by the night sky brightness and detector cost. Concepts with alternative semiconductors \citep{Sullivan2014} or
creative optics on polar sites (Moore et al. in prep) are being investigated. If WFIRST elects a suitable survey design and prioritizes near real-time transient 
identification, it could be a powerful probe to discover IR transients.  

In summary, the SPIRITS discovery of SPRITEs bodes well for future wide-field explorations of the dynamic 
IR sky.

\bigskip
\bigskip
We thank O. Pejcha, E. Lovegrove, S. Woosley, A. L. Piro, E. S. Phinney, S. R. Kulkarni, L. Bildsten and E. Quataert for valuable discussions. 
This work is based on observations made with the Spitzer Space Telescope, which is operated by the Jet Propulsion Laboratory, California Institute of Technology under a contract with NASA. 
%Support for this work was provided by NASA through an award issued by JPL/Caltech. 
The SPIRITS team acknowledges generous support from the NASA {\it Spitzer} grants for SPIRITS. 
MMK thanks the National Science Foundation for a PIRE Grant No. 1545949 for the GROWTH project. JJ acknowledges the National Science Foundation Graduate Research Fellowship under 
Grant No. DGE-1144469. PAW and SM are grateful to the South African National Research Foundation (NRF) for a research grant.
RDG and the MLOF group were supported, in part, by the United Stated Air Force.

%\begin{figure*}[!hbt]
%\centering
%\includegraphics[height=0.3\textwidth]{SPIRITS14aje.png}\includegraphics[height=0.27\textwidth]{14aje_host.png}
%\caption{ {\it Left: } Discovery image of a new IR transient SPIRITS\,14\,aje in Messier\,101. {\it Left Middle:} Reference image from Archival {\it Spitzer} Data. {\it Right Middle:} Difference image. {\it Right:} Location on spiral arm in M\,101. 
%%The mid-IR absolute magnitude is $-$12.8\,mag. 
%\label{fig:14aje}}
%\end{figure*}

%SNAPSHOTS from SPIRITS Marshal stored in khagol:~/papers/spirits/thumbs/

%CODE: khagol:~/papers/spirits/hosts/ %%ZOOM=0.8, tile with 3x5, gap=1pix, scale=minmax for last three and zscale for others.

%CODE: khagol:~/papers/spirits/giros/lum_color.pro and ~/papers/spirits/giros/lc_collage_sprites.pro (Note:latest version is on laptop)

%CODE: khagol:~/papers/spirits/giros/plotlc.pro

%CODE: khagol:~/papers/spirits/giros/plotlc.pro
\begin{figure*}[!hbt] 
\centering
\includegraphics[width=\textwidth, height=\textwidth]{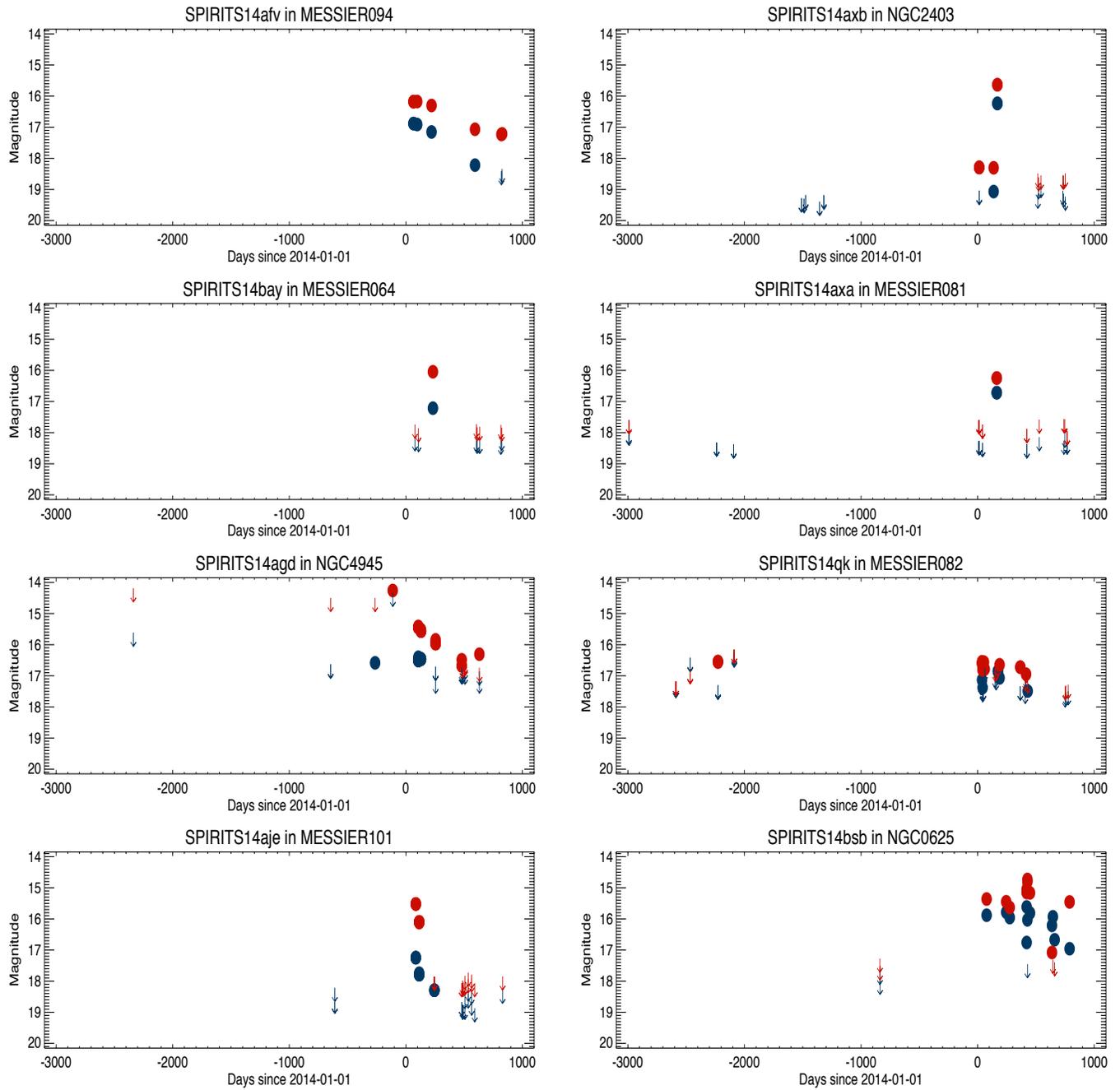}
\caption{IR light curves of relatively fast evolving SPRITEs. }
\label{fig:slowlc}
\end{figure*}

%CODE: khagol:~/papers/spirits/14ajc - made in ds9
\begin{figure*}[!hbt] 
\centering
\includegraphics[width=1.2\textwidth,angle=270]{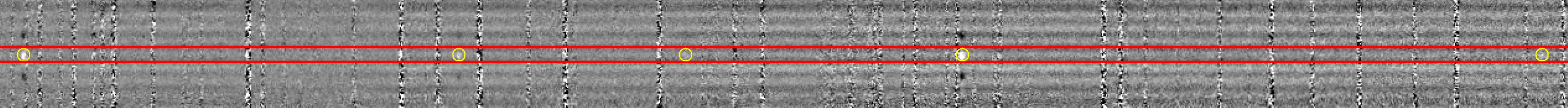}
\caption{Calibrated and rectified 2-d spectrum of SPIRITS\,14ajc (y-axis is the spectral and x-axis is the spatial dimension). Note the
five lines of excited molecular hydrogen (yellow circles) and lack of continuum at the trace position of SPIRITS\,14jac (red lines). The spectrum
spans 19565\AA\, to 22320\AA\, in the K-band. The horizontal striped lines are sky line residuals.}
\label{fig:slowlc}
\end{figure*}

\begin{deluxetable*}{lllll}
  \tabletypesize{\scriptsize}
  \tablecaption{SPRITEs: Co-ordinates and Host Galaxies \label{tab:radec}}
  \tablewidth{0pt}
  \tablehead{\colhead{Name} & \colhead{RA (J2000)} & \colhead{DEC (J2000)} &  \colhead{Host Galaxy} & \colhead{Distance Modulus}}
  \startdata
SPIRITS\,14qk  & $\RA{09}{55}{28.72}$   & $+\DEC{69}{39}{58.6}$  & MESSIER\,82 & 27.73 \citep{2009AJ....138..332J} \\  %Very faint progenitor, Spiral Host
SPIRITS\,14afv & $\RA{12}{50}{49.56}$   & $+\DEC{41}{05}{52.7}$  & MESSIER\,94 & 28.31 \citep{2013AJ....146...86T}   \\  %faint progenitor, Spiral Host
SPIRITS\,14agd & $\RA{13}{05}{30.87}$   & $-\DEC{49}{26}{50.8}$  & NGC\,4945    & 27.90 \citep{2008ApJ...686L..75M}   \\  %no progenitor, spiral not in sdss
SPIRITS\,14ajc  & $\RA{13}{36}{52.95}$   & $-\DEC{29}{52}{16.1}$  & MESSIER\,83 & 28.41  \citep{2011ApJS..195...18R}  \\  %no progenitor, spiral not in sdss
SPIRITS\,14ajd & $\RA{13}{37}{05.02}$   & $-\DEC{29}{48}{56.2}$  & MESSIER\,83 & 28.41  \citep{2011ApJS..195...18R}  \\  %no progenitor, spiral not in sdss
SPIRITS\,14aje & $\RA{14}{02}{55.51}$   & $+\DEC{54}{23}{18.5}$  & MESSIER\,101 & 29.34  \citep{2015AstL...41..239T} \\  %no progenitor, spiral, iPTF limit
SPIRITS\,14ajp & $\RA{13}{37}{12.71}$   & $-\DEC{29}{49}{14.9}$  & MESSIER\,83 & 28.41  \citep{2011ApJS..195...18R}  \\ %faint progenitor, spiral not in sdss
SPIRITS\,14ajr  & $\RA{13}{36}{54.81}$   & $-\DEC{29}{52}{33.7}$  & MESSIER\,83 & 28.41  \citep{2011ApJS..195...18R}  \\  %no progenitor, spiral not in sdss
SPIRITS\,14ave & $\RA{03}{47}{03.17}$   & $+\DEC{68}{09}{05.3}$  & IC\,342 & 27.58  \citep{2014AJ....148....7W}      \\ %no progenitor, outskirts of spiral not in sdss $-$$-$ no IPTF Data 
SPIRITS\,14axa & $\RA{09}{56}{01.52}$   & $+\DEC{69}{03}{12.5}$  & MESSIER\,81 & 27.80   \citep{2012ApJ...751L..19J} \\ %no progenitor, spiral, iPTF limit
SPIRITS\,14axb & $\RA{07}{36}{34.70}$   & $+\DEC{65}{39}{22.4}$  & NGC\,2403 & 27.51  \citep{2011ApJS..195...18R}     \\  %faint progenitor, spiral, iPTF limit
SPIRITS\,14bay & $\RA{12}{56}{43.25}$   & $+\DEC{21}{42}{25.7}$  & MESSIER\,64 & 29.37  \citep{2013AJ....146...86T}  \\  %very faint progenitor, spiral
SPIRITS\,14bgq & $\RA{13}{39}{50.99}$   & $-\DEC{31}{38}{46.0}$  & NGC\,5253 & 27.76  \citep{2008ApJ...686L..75M}     \\ %faint progenitor, spiral not in sdss
%SPIRITS\,14bmc & $\RA{00}{54}{49.68}$   & $-\DEC{37}{39}{51.2}$  & NGC\,300 & 26.44  \citep{2009ApJS..183...67D}     \\  %no progenitor, spiral not in sdss
SPIRITS\,14bsb & $\RA{01}{35}{06.72}$   & $-\DEC{41}{26}{13.5}$  & NGC\,625 & 28.12   \citep{2009AJ....138..332J}    \\ %no progenitor, spiral not in sdss   
\enddata
\end{deluxetable*}

\begin{deluxetable*}{cccccccccccccc}
  \tabletypesize{\scriptsize}
  \tablecaption{Light Curve Properties of SPRITEs \label{tab:lcprop}}
  \tablewidth{0pt}
  \tablehead{\colhead{Name} & \colhead{Peak Time} & \colhead{Peak Mag} & \colhead{Peak Time} & \colhead{Peak Mag} & \colhead{Color Peak Time} & \colhead{Color Peak Mag} & \colhead{Progenitor limit}  & \colhead{Progenitor limit} & \colhead{Lifespan} & \colhead{Evolution} & \colhead{Lifespan} & \colhead{Evolution} & \colhead{Speed} \\
   & \colhead{[3.6]} & \colhead{[3.6]} & \colhead{[4.5]} & \colhead{[4.5]} & \colhead{[3.6]-[4.5]} & \colhead{[3.6]-[4.5]} & \colhead{[3.6]} & \colhead{[4.5]} & \colhead{[3.6]} & \colhead{[3.6]} & \colhead{[4.5]} & \colhead{[4.5]} &  \\
   &  \colhead{MJD}  & \colhead{Mag}  & \colhead{MJD}   &  \colhead{Mag}  & \colhead{MJD}         & \colhead{Mag}         & \colhead{Mag}   & \colhead{Mag}   & \colhead{yr}   & \colhead{mag yr$^{-1}$} & \colhead{yr} & \colhead{mag yr$^{-1}$} & 
  }
  \startdata
SPIRITS\,14ave & 57007.4 & $-$11.2 & 57007.4 & $-$11.8 & 57182.2 &  0.3 & $-$10.6   & $-$10.8   &  1.0 &   0.1 &  1.0 &   0.5 & Slow \\
SPIRITS\,14afv & 56722.7 & $-$11.4 & 56754.9 & $-$12.1 & 57250.7 &  1.2 & $-$10.6   & $-$10.9   &  1.4 &   0.9 &  2.1 &   0.5 & Fast \\
SPIRITS\,14ajp & 56787.2 & $-$11.3 & 55290.7 & $-$11.9 & 56915.5 &  0.7 & $-$10.0   & $-$10.3   &  5.6 &   0.2 &  4.4 &   $<$0.1 & Slow \\
SPIRITS\,14axb & 56827.4 & $-$11.3 & 56827.4 & $-$11.9 & 56827.4 &  0.6 & $-$7.90   & $-$9.37   &  0.1 & $-$32.2 &  0.4 &  $-$6.2 & Fast \\
SPIRITS\,14bay & 56889.8 & $-$11.4 & 56889.8 & $-$12.6 & 56889.8 &  1.2 & $-$11.4   & $-$11.5   &  \nodata &  \nodata &  \nodata &   \nodata & Fast \\
SPIRITS\,14ajr & 56915.5 & $-$11.3 & 56787.2 & $-$12.2 & 56915.5 &  0.8 & $-$11.1   & $-$11.5   &  4.4 &  $-$0.1 &  5.0 &   0.1 & Slow \\
SPIRITS\,14axa & 56821.9 & $-$11.1 & 56821.9 & $-$11.6 & 56821.9 &  0.5 & $-$9.19   & $-$9.56   &  \nodata &   \nodata &  \nodata &   \nodata & Fast \\
SPIRITS\,14agd & 56764.5 & $-$11.5 & 56544.9 & $-$13.6 & 56787.1 &  0.9 & $-$11.3   & $-$11.7   &  1.1 &  $-$0.1 &  1.6 &   1.5 & Fast \\
SPIRITS\,14bgq & 55424.4 & $-$10.8 & 55424.4 & $-$11.6 & 57312.2 &  0.8 & \nodata & \nodata &  5.6 &  $<$0.1 &  5.6 &  $<$0.1 & Slow \\
SPIRITS\,14qk  & 56831.7 & $-$10.9 & 54430.7 & $-$11.2 & 56846.3 &  0.4 & $-$10.6   & $-$10.7   &  0.7 &   0.9 &  7.1 &   $<$0.1 & Fast \\
SPIRITS\,14ajc & 56765.1 & $-$11.3 & 56765.1 & $-$12.2 & 57161.2 &  0.5 & $-$10.6   & $-$11.4   &  5.1 &   $<$0.1 &  5.1 &   0.1 & Slow \\
SPIRITS\,14aje & 56742.8 & $-$11.9 & 56742.8 & $-$13.7 & 56771.8 &  1.6 & $-$10.5   & $-$11.6   &  0.4 &   2.4 &  0.1 &   7.3 & Fast \\
%SPIRITS\,14bmc & 56905.2 & $-$10.7 & 56933.4 & $-$11.6 & 56933.4 &  1.0 & -8.45   & -8.83   &  0.1 &   1.3 &  0.1 &  -4.4 & Fast \\
SPIRITS\,14ajd & 55290.7 & $-$11.2 & 55290.7 & $-$12.2 & 56915.5 &  1.0 & $-$10.3   & $-$10.9   &  4.4 &   0.1 &  5.1 &   0.3 & Slow \\
SPIRITS\,14bsb & 57078.5 & $-$12.5 & 57085.0 & $-$13.3 & 57446.6 &  1.5 & \nodata & \nodata &  1.9 &   0.6 &  1.9 &   $<$0.1 & Fast \\
\enddata
\end{deluxetable*}

\begin{deluxetable*}{cccccccccc}
  \tabletypesize{\scriptsize}
  \tablecaption{SPIRITS\,14ajc Emission Lines \label{tab:14ajc}}
  \tablewidth{0pt}
  \tablehead{\colhead{Transition} & \colhead{Wavelength} & \colhead{Flux} &  \colhead{GFWHM} & \colhead{Relative Velocity} & \colhead{Flux Ratio} & \colhead{Flux Ratio} & \colhead{Flux Ratio} & \colhead{Flux Ratio} & \colhead{Flux Ratio}  \\
 & & & & & (Observed)  & \colhead{if 1000K} & \colhead{if 2000K} & \colhead{if 3000K} & \colhead{if 4000K}}
  \startdata
1-0 S(0) & 22275.1 &  11.43 & 8.96 & 28    &  0.28 &  0.27 &  0.21 &  0.19 &  0.19 \\   
1-0 S(1) & 21258.9 &  40.13 & 6.74 & 65    &  1.00 &  1.00 &  1.00 &  1.00 &  1.00 \\ 
2-1 S(3) & 20774.4 &  8.013 & 9.74 & 57    &  0.20 &  0.003&  0.084&  0.27 &  0.47 \\ 
1-0 S(2) & 20376.4 &  10.61 & 6.04 & 53    &  0.26 &  0.27 &  0.37 &  0.42 &  0.44 \\ 
1-0 S(3) & 19613.6 &  25.67 & 6.71 & 63    &  0.64 &  0.51 &  1.02 &  1.29 &  1.45 \\
 \enddata
\end{deluxetable*}

\begin{deluxetable*}{ccccccccc}
  \tabletypesize{\scriptsize}
  \tablecaption{Follow-up Photometry \label{tab:followup}}
  \tablewidth{0pt}
  \tablehead{\colhead{Name} &  \colhead{Observation Date} & \colhead{Telescope} & \colhead{Instrument} & \colhead{Filter} & \colhead{Photometry} }
  \startdata
SPIRITS\,14qk & 2014 stack (N=74) & P48 & iPTF & R & $>$23.5 \\ 
SPIRITS\,14afv & 2014 stack (N=117) & P48 & iPTF & R  & $>$20.1 \\ % g > 20.1
      & 2014-06-19  & MLOF  & 2MASS     & J  & $>$17.4 \\
      & 2014-06-19  & MLOF  & 2MASS     & H  & $>$17.3 \\
      & 2014-06-19  & MLOF  & 2MASS     & Ks  & $>$16.2 \\
SPIRITS\,14agd & 2014 stack (N=16) & Swope & CCD & g & $>$21.7 \\ 
      & 2014 stack (N=16) & Swope & CCD & r & $>$21.2 \\ 
      & 2014 stack (N=16) & Swope & CCD & i & $>$21.0 \\ 
SPIRITS\,14ajc & 2014-04-20  & Swope & CCD     & g & $>$20.0 \\
      & 2014-04-20  & Swope & CCD    & r & $>$20.0 \\
      & 2014-04-20  & Swope & CCD    & i & $>$19.8 \\ 
%      & 2014-05-24  & MLOF  & 2MASS    & J  & $>$15.0 \\ %17.3 \\
%     & 2014-05-24  & MLOF  & 2MASS     & H  & $>$14.4 \\ %16.8 \\   
%     & 2014-05-24  & MLOF  & 2MASS     & Ks  & $>$13.9 \\ %15.7 \\  
%      & 2014-05-18  & du Pont & Retrocam & Y & ?? \\
      & 2014-05-18  & du Pont & Retrocam & J & $>$19.8 \\
      & 2014-05-18  & du Pont & Retrocam & H  & $>$18.7 \\ 
      & 2014-07-02  & Keck  & DEIMOS & I  & $>$23.5 \\
      & 2014-06-07  & Keck   & MOSFIRE & Ks & $>$18.7 \\
      & 2012-09-03  & HST   & WFC3   & H  & $>$ 22.0 \\
      & 2012-07-22  & HST   & WFC3   & I  & $>$ 25.5 \\
SPIRITS\,14ajd  & 2014-04-20  & Swope & CCD     & g & $>$20.0 \\
      & 2014-04-20  & Swope & CCD    & r & $>$20.0 \\
      & 2014-04-20  & Swope & CCD    & i & $>$19.8 \\ 
      & 2014-06-07  & Keck   & MOSFIRE & Ks & $>$19.6 \\
      & 2014-05-24  & MLOF  & 2MASS     & J  & $>$17.3 \\
      & 2014-05-24  & MLOF  & 2MASS     & H  & $>$16.5 \\
      & 2014-05-24 & MLOF  & 2MASS     & Ks  & $>$16.4 \\      
SPIRITS\,14aje & 2014 stack (N=157) & P48 & iPTF & R & $>$23.9 \\ % g > 20.9
      & 2014-05-01  & MLOF  & 2MASS     & J  & $>$18.3 \\
      & 2014-05-01  & MLOF  & 2MASS     & H  & $>$16.8 \\
      & 2014-05-01  & MLOF  & 2MASS     & Ks  & $>$16.4 \\
      & 2014-07-02  & Keck  & DEIMOS & I  & $>$24.3 \\
      & 2014-06-07  & Keck   & MOSFIRE & Ks & $>$19.4 \\  
      & 2014-06-07  & Keck   & MOSFIRE & J & $>$20.5 \\
      & 2014-09-23  & HST & WFC3 & I  & $>$26.5 \\
      & 2014-09-23  & HST & WFC3 & J & $>$25.0 \\
      & 2014-09-23  & HST & WFC3 & H & $>$22.5  \\    
SPIRITS\,14ajp  & 2014-04-20  & Swope & CCD     & g & $>$20.0 \\
      & 2014-04-20  & Swope & CCD    & r & $>$20.0 \\
      & 2014-04-20  & Swope & CCD    & i & $>$19.8 \\ 
      & 2014-06-07  & Keck   & MOSFIRE & Ks & $>$19.2 \\
%      & 2014-05-24  & MLOF  & 2MASS  & J  & $>$13.3 \\
 %     & 2014-05-24  & MLOF  & 2MASS     & H  & $>$12.8 \\
 %     & 2014-05-24 & MLOF  & 2MASS     & Ks  & $>$13.1 \\    
SPIRITS\,14ajr  & 2014-04-20  & Swope & CCD     & g & $>$20.0 \\
      & 2014-04-20  & Swope & CCD    & r & $>$20.0 \\
      & 2014-04-20  & Swope & CCD    & i & $>$19.8 \\ 
      & 2014-06-07  & Keck   & MOSFIRE & Ks & $>$19.2 \\
      & 2014-05-24  & MLOF  & 2MASS     & J  & $>$17.3 \\
      & 2014-05-24  & MLOF  & 2MASS     & H  & $>$16.5 \\
      & 2014-05-24 & MLOF  & 2MASS     & Ks  & $>$16.4 \\ 
SPIRITS\,14ave & 2014-09-23 & LCOGT-1m & SBIG & i  & $>$20.0 \\     
& 2014-10-26 & LCOGT-1m & SBIG & i  & $>$20.3 \\                         
SPIRITS\,14axa & 2014 stack (N=117) & P48 & iPTF & R  & $>$23.15 \\
                         %& 2014-09-26             & HST & WFC3 & I  & $>$26.5?  \\
                         %& 2014-09-26             & HST & WFC3 & J & $>$ \\
                         & 2014-09-26             & HST & WFC3 & H & $>$22.5 \\
SPIRITS\,14axb & 2014 stack (N=18) & P48 & iPTF & R & $>$22.5 \\
     & 2014-05-01 & MLOF & 2MASS & J &  $>$14.2 \\
    & 2014-05-01 & MLOF & 2MASS & H &  $>$14.0 \\
    & 2014-05-01 & MLOF & 2MASS & Ks &  $>$13.3 \\
SPIRITS\,14bay & 2014 stack (N=66) & P48 & iPTF & R & $>$23.3 \\
SPIRITS\,14bgq & 2014 stack (N=12) & Swope & CCD & g & $>$21.9 \\ 
     & 2014 stack (N=12) & Swope & CCD & r & $>$20.9 \\ 
     & 2014 stack (N=10) & Swope & CCD & i & $>$21.0 \\
      & 2014-04-30  & MLOF  & 2MASS     & J  & $>$16.9 \\
      & 2014-04-30  & MLOF  & 2MASS     & H  & $>$15.6 \\
      & 2014-04-30  & MLOF  & 2MASS     & Ks  & $>$15.7 \\
%SPIRITS\,14bmc & 2014 stack (N=13) & Swope & CCD & g & $>$22.4 \\ 
%     & 2014 stack (N=14) & Swope & CCD & r & $>$21.8 \\ 
%     & 2014 stack (N=15) & Swope & CCD & i & $>$21.7 \\  
%      & 2014-11-23  & MLOF  & 2MASS     & J  & $>$16.0 \\
%      & 2014-11-23  & MLOF  & 2MASS     & H  & $>$16.0 \\
%      & 2014-11-23  & MLOF  & 2MASS     & Ks  & $>$15.3 \\
SPIRITS\,14bsb & 2014 stack (N=13) & Swope & CCD & g & $>$22.9 \\ 
     & 2014 stack (N=13) & Swope & CCD & r & $>$22.6 \\ 
     & 2014 stack (N=13) & Swope & CCD & i & $>$22.1 \\
  \enddata
   \tablecomments{This Table is published in its entirety (including individual epochal observations) in the machine-readable format. Upper limits are 5$\sigma$.}
\end{deluxetable*}

%\begin{longtable}
\begin{deluxetable*}{cccc}
  \tabletypesize{\scriptsize}
  \tablecaption{Spitzer Photometry of all SPRITEs \label{tab:allphot}}
  \tablewidth{0pt}
  \tablehead{\colhead{Name} &  \colhead{Observation Date} & \colhead{Filter} & \colhead{Magnitude} }
  \startdata
SPIRITS\,14aje & 53072.09 & [3.6] & $>$19.25 \\ 
SPIRITS\,14aje & 53072.49 & [3.6] & $>$17.93 \\ 
SPIRITS\,14aje & 56048.39 & [3.6] & $>$18.59 \\ 
SPIRITS\,14aje & 56742.84 & [3.6] & 17.25 $\pm$  0.01 \\ 
SPIRITS\,14aje & 56771.83 & [3.6] & 17.79 $\pm$  0.02 \\ 
SPIRITS\,14aje & 56771.83 & [3.6] & 17.74 $\pm$  0.01 \\ 
SPIRITS\,14aje & 56902.01 & [3.6] & 18.29 $\pm$  0.02 \\ 
SPIRITS\,14aje & 57136.69 & [3.6] & $>$18.68 \\ 
SPIRITS\,14aje & 57144.06 & [3.6] & $>$18.79 \\ 
SPIRITS\,14aje & 57150.17 & [3.6] & $>$18.77 \\ 
SPIRITS\,14aje & 57163.72 & [3.6] & $>$18.79 \\ 
SPIRITS\,14aje & 57163.72 & [3.6] & $>$18.44 \\ 
SPIRITS\,14aje & 57191.83 & [3.6] & $>$18.43 \\ 
SPIRITS\,14aje & 57191.83 & [3.6] & $>$18.20 \\ 
SPIRITS\,14aje & 57220.80 & [3.6] & $>$18.66 \\ 
SPIRITS\,14aje & 57247.82 & [3.6] & $>$18.87 \\ 
SPIRITS\,14aje & 57486.85 & [3.6] & $>$18.27 \\ 
SPIRITS\,14aje & 56742.84 & [4.5] & 15.52 $\pm$  0.01 \\ 
SPIRITS\,14aje & 56771.83 & [4.5] & 16.10 $\pm$  0.01 \\ 
SPIRITS\,14aje & 56771.83 & [4.5] & 16.12 $\pm$  0.01 \\ 
SPIRITS\,14aje & 56902.01 & [4.5] & $>$17.85 \\ 
SPIRITS\,14aje & 57136.69 & [4.5] & $>$18.06 \\ 
SPIRITS\,14aje & 57144.06 & [4.5] & $>$17.98 \\ 
SPIRITS\,14aje & 57150.17 & [4.5] & $>$18.05 \\ 
SPIRITS\,14aje & 57163.72 & [4.5] & $>$18.01 \\ 
SPIRITS\,14aje & 57163.72 & [4.5] & $>$17.83 \\ 
SPIRITS\,14aje & 57191.83 & [4.5] & $>$17.94 \\ 
SPIRITS\,14aje & 57191.83 & [4.5] & $>$17.73 \\ 
SPIRITS\,14aje & 57220.80 & [4.5] & $>$17.91 \\ 
SPIRITS\,14aje & 57247.82 & [4.5] & $>$18.06 \\ 
SPIRITS\,14aje & 57486.85 & [4.5] & $>$17.85 \\ 
\hline
SPIRITS\,14axb & 53286.42 & [3.6] & $>$19.53 \\ 
SPIRITS\,14axb & 55146.68 & [3.6] & $>$19.28 \\ 
SPIRITS\,14axb & 55168.73 & [3.6] & $>$19.31 \\ 
SPIRITS\,14axb & 55181.97 & [3.6] & $>$19.20 \\ 
SPIRITS\,14axb & 55302.72 & [3.6] & $>$19.39 \\ 
SPIRITS\,14axb & 55339.45 & [3.6] & $>$19.20 \\ 
SPIRITS\,14axb & 56671.43 & [3.6] & $>$19.04 \\ 
SPIRITS\,14axb & 56795.32 & [3.6] & 19.07 $\pm$  0.03 \\ 
SPIRITS\,14axb & 56827.40 & [3.6] & 16.24 $\pm$  0.01 \\ 
SPIRITS\,14axb & 57174.60 & [3.6] & $>$19.16 \\ 
SPIRITS\,14axb & 57180.81 & [3.6] & $>$18.86 \\ 
SPIRITS\,14axb & 57201.79 & [3.6] & $>$18.82 \\ 
SPIRITS\,14axb & 57388.85 & [3.6] & $>$19.04 \\ 
SPIRITS\,14axb & 57395.55 & [3.6] & $>$19.13 \\ 
SPIRITS\,14axb & 57409.70 & [3.6] & $>$19.23 \\ 
SPIRITS\,14axb & 56671.43 & [4.5] & 18.30 $\pm$  0.03 \\ 
SPIRITS\,14axb & 56795.32 & [4.5] & 18.30 $\pm$  0.03 \\ 
SPIRITS\,14axb & 56827.40 & [4.5] & 15.64 $\pm$  0.01 \\ 
SPIRITS\,14axb & 57174.60 & [4.5] & $>$18.48 \\ 
SPIRITS\,14axb & 57180.81 & [4.5] & $>$18.61 \\ 
SPIRITS\,14axb & 57201.79 & [4.5] & $>$18.55 \\ 
SPIRITS\,14axb & 57388.85 & [4.5] & $>$18.54 \\ 
SPIRITS\,14axb & 57395.55 & [4.5] & $>$18.55 \\ 
SPIRITS\,14axb & 57409.70 & [4.5] & $>$18.48 \\ 
  \enddata
  \tablecomments{This Table is published in its entirety (including all SPRITEs) in the machine-readable format.
      A portion is shown here for guidance regarding its form and content.}
  \end{deluxetable*}

\end{document}